\documentclass[usenatbib,useAMS]{mn2e}
\usepackage[titletoc,toc]{appendix}
\usepackage{epsfig,psfig,graphicx,float,color}
\usepackage{subfigure}
\usepackage{mybibdefs}
\usepackage{multirow}
\usepackage{widetext}
\usepackage{url}
 \usepackage{color}
\usepackage{amsmath}

\newcommand\rr{\color{black}}

\newcommand{\bra}{\langle}
\newcommand{\ket}{\rangle}

\usepackage{graphicx}
\usepackage{amssymb}
\usepackage{epstopdf}
\title[Anomaly detection for machine learning redshifts]{Anomaly detection for machine learning redshifts applied to SDSS galaxies}  
\author[Hoyle et al.]{Ben  Hoyle$^{1,2}$, Markus Michael Rau$^{1,4}$, Kerstin Paech$^{1,2}$, Christopher Bonnett$^3$  \newauthor Stella Seitz$^{1,4}$, Jochen Weller$^{1,2,4}$\\\\
$^1$Universitaets-Sternwarte, Fakultaet fuer Physik, Ludwig-Maximilians Universitaet Muenchen, Scheinerstr. 1, D-81679 Muenchen, Germany\\
$^2$Excellence Cluster Universe, Boltzmannstr. 2, D-85748 Garching, Germany\\
$^3$Institut de Fõsica d'Altes Energies, Universitat Autonoma de Barcelona, E-08193 Bellaterra, Spain\\
$^4$Max Planck Institute for Extraterrestrial Physics, Giessenbachstr. 1, D-85748 Garching, Germany\\
\\
{\tt E-mail: hoyleb@usm.uni-muenchen.de}
} 
  
\begin{document}
\date{Accepted ----. Received ----; in original form ----.}
\pagerange{\pageref{firstpage}--\pageref{lastpage}} \pubyear{2010}
\maketitle
\label{firstpage}
\begin{abstract}
We present an analysis of anomaly detection for machine learning redshift estimation. Anomaly detection allows the removal of poor training examples, which can adversely influence redshift estimates. Anomalous training examples may be photometric galaxies with incorrect spectroscopic redshifts, or  galaxies with one or more poorly measured photometric quantity.

We select 2.5 million `clean' SDSS DR12 galaxies with reliable spectroscopic redshifts, and 6730 `anomalous' galaxies with spectroscopic redshift measurements which are flagged as unreliable.
We contaminate the clean base galaxy sample with galaxies with unreliable redshifts and attempt to recover the contaminating galaxies using the Elliptical Envelope technique. We then train four machine learning architectures for redshift analysis on both the contaminated sample and on the preprocessed `anomaly-removed' sample and measure redshift statistics on a clean validation sample generated without any preprocessing.  We find an improvement on all measured statistics of up to $80\%$  when training on the anomaly removed sample as compared with training on the contaminated sample for each of the machine learning routines explored.  We further describe a method to estimate the contamination fraction of a base data sample.
\end{abstract}
\begin{keywords}
galaxies: distances and redshifts,  catalogues, surveys.
\end{keywords}

\section{introduction}
Photometric surveys can be maximally exploited for large scale structure analyses once galaxies have been identified and their positions on the sky and in redshift space have been measured. Measuring accurate spectroscopic redshifts is costly and time intensive, and is typically only performed for a small subsample of all galaxies. For this subsample of galaxies one may learn a mapping between the measured photometric properties, and the spectroscopic redshift. This mapping can then be applied to all photometrically identified galaxies to estimate redshifts. This is the basis of machine learning, and inherently assumes that the galaxies used to construct the mapping form an unbiased and uncontaminated sample of the final dataset. 

Recent work by the current authors shows that if the base training sample is biased compared to the final sample, it may be augmented, e.g., by adding galaxies from simulations, to make the data sets appear more similar \citep[][]{2015MNRAS.450..305H}. The data augmentation process has been shown to improve the redshift estimate of the final test sample. In this paper we examine the problem of identifying poorly measured galaxy properties which contaminate the base training set. The contamination may be due to incorrectly measured spectroscopic redshifts, or unreliable photometric properties. 

Photometric redshifts are also estimated by parametric techniques, for example from galaxy Spectral Energy Distribution (hereafter SED) templates. Some templates encode our knowledge of stellar population models which result in predictions for the evolution of galaxy magnitudes and colors. This parametric encoding of the complex stellar physics coupled with the uncertainty of the parameters of the stellar population models combine to produce redshift estimates which are little better than many non-parametric techniques \citep[see e.g.,][for an overview of different techniques]{2010A&A...523A..31H,2013ApJ...775...93D}. 

When a representative training sample is available, machine learning methods offer an alternative to template methods to estimate galaxy redshifts. The `machine architecture' determines how to best manipulate the photometric galaxy input properties (or `features') to produce a machine learning redshift. The machine attempts to learn the most effective manipulations to minimize the difference between the spectroscopic redshift and the machine learning redshift of the training sample.

The field of machine learning for photometric redshift analysis has been developing since \cite{2003LNCS.2859..226T} used artificial Neural Networks (aNNs). A plethora of machine learning architectures, including tree based methods, have been 
applied to the problem of point prediction redshift estimation \citep[see e.g.][for a further list and routine comparisons]{2014MNRAS.445.1482S}, or to estimate the full redshift probability distribution function \citep[hereafter pdf,][]{2010ApJ...715..823G,tpz,2015MNRAS.449.1043B,2015arXiv150308215R}. Machine learning architectures have also had success in other fields of astronomy 
such as galaxy morphology identification, and star \& quasar separation \citep[see for example][]{1997daa..conf...43L,2009arXiv0910.3770Y}.

It is often assumed that the training data does not contain galaxies with unreliable spectroscopic redshift estimates, or does not contain galaxies with incorrectly estimated photometric properties. However the contamination of a training sample can adversely affect the recovered machine learning redshifts. The authors \cite{2014MNRAS.444..129C} use simulated spectra to show how the cosmological constraints for a weak lensing survey are degraded in the presence of even 1\% of spectroscopic outliers in the training sample. 

Previous work on outlier analysis has been confined to examining the properties of the machine learning redshifts {\it after} the system has been trained. For example, photometric redshift `outliers' which actually sit in a different redshift bin than expected, can be identified by cross correlating data across bins \citep[see e.g.,][]{2006ApJ...651...14S,2010MNRAS.401.1399B,2013MNRAS.433.2857M}. Training data can also be carefully removed if the final machine learning redshift and the spectroscopic redshift are found to be very dissimilar \citep[][]{2014MNRAS.444..129C}. More recently outlier detection has been performed on galaxies after a pdf has been obtained, by examining the pdf for multiple peaks or other irregularities \citep[][]{2014MNRAS.442.3380C}. All of these methods enable the construction of a cleaner final sample of galaxies. However the cleaned sample must first be carefully checked to ensure that a sample bias has not been introduced before being used for scientific analysis. In particular the final test sample must be made to be representative of the cleaned training sample.

In this paper we explore the effect of performing outlier analysis, or anomaly detection, on the training sample to identify discrepant photometric data, or unreliable spectroscopic redshifts {\it before} the sample is used to estimate a machine learning redshift. We then show how the removal of this anomalous data improves the machine learning redshift metrics for two very different groups of machine learning architectures.

This paper is organized as follows: In \S\ref{data} we describe the data sample and the machine learning methods employed; We present the anomaly analysis and improvement to the redshift estimates using the anomaly detection in \S\ref{results};  and conclude  in \S\ref{conclusions}.

\section{Data  and Machine Architectures}
\label{data}
In this study we use observational data drawn from the final SDSS data release, and explore a selection of machine learning architectures for anomaly detection and machine learning redshift estimation.

\subsection{Observational dataset}
\label{obs_data}
The observational data in this study are drawn from SDSS III Data Release 12 \citep[][]{2015arXiv150100963A}. The SDSS I-III uses a 4 meter telescope at Apache Point Observatory in New Mexico and has CCD wide field photometry in 5 bands \citep[$u,g,r,i,z$][]{Gunn:2006tw,Smith:2002pca}, and an expansive spectroscopic follow up program \citep[][]{2011AJ....142...72E} covering $\pi$ steradians of the northern sky. The SDSS collaboration has obtained more than 3 million galaxy spectra using dual fiber-fed spectrographs. An automated photometric pipeline performs object classification to a magnitude of $r\approx22$ and measures photometric properties of more than 100 million galaxies. The complete data sample, and many derived catalogs such as the photometric properties, are publicly available through the CasJobs server \citep[][]{10.1109/MCSE.2008.6}\footnote{skyserver.sdss3.org/CasJobs}.

\begin{figure}
 \includegraphics[scale=0.47,clip=true,trim=0 15 40 30]{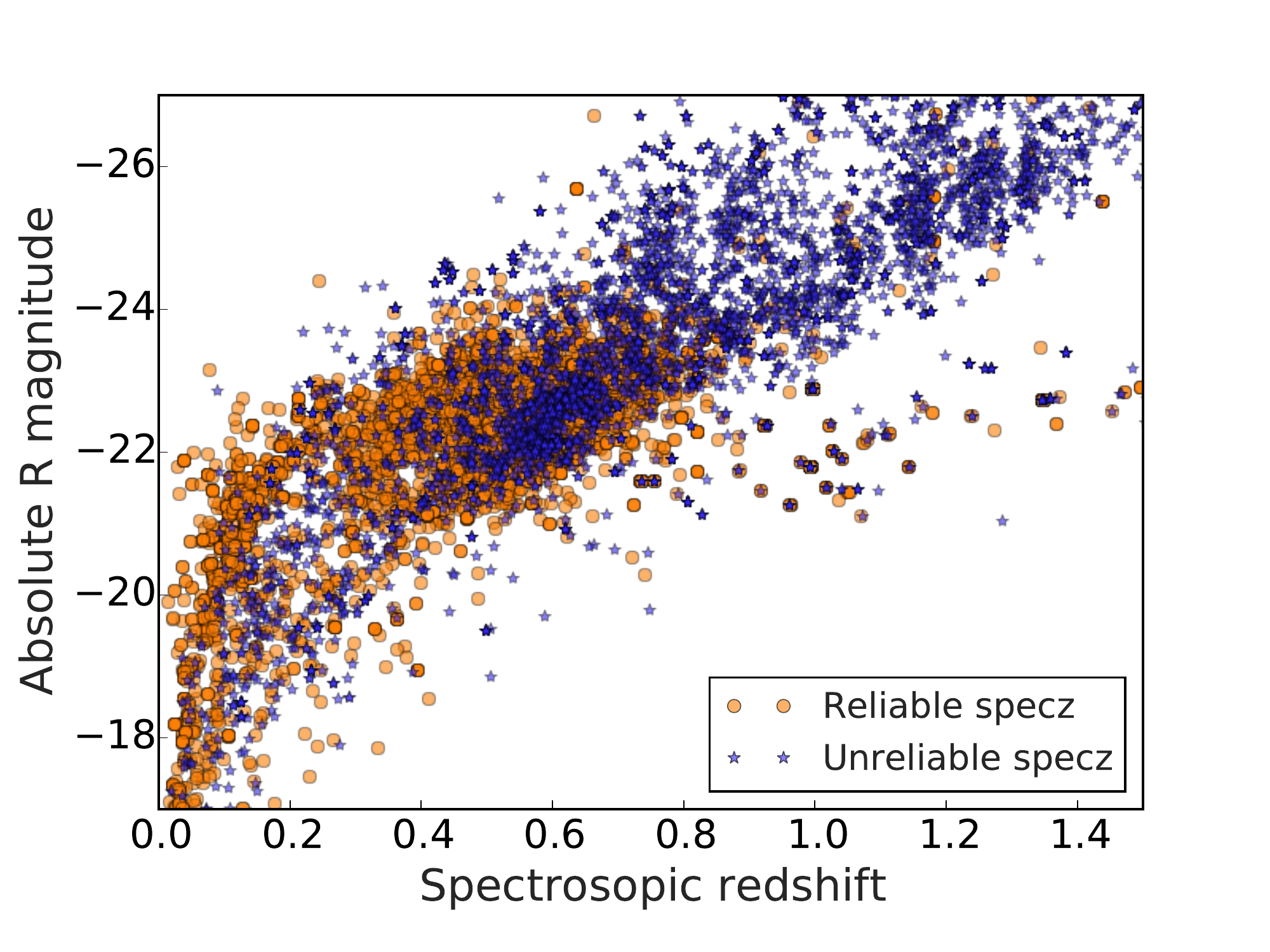}
 \includegraphics[scale=0.47,clip=true,trim=0 15 40 30]{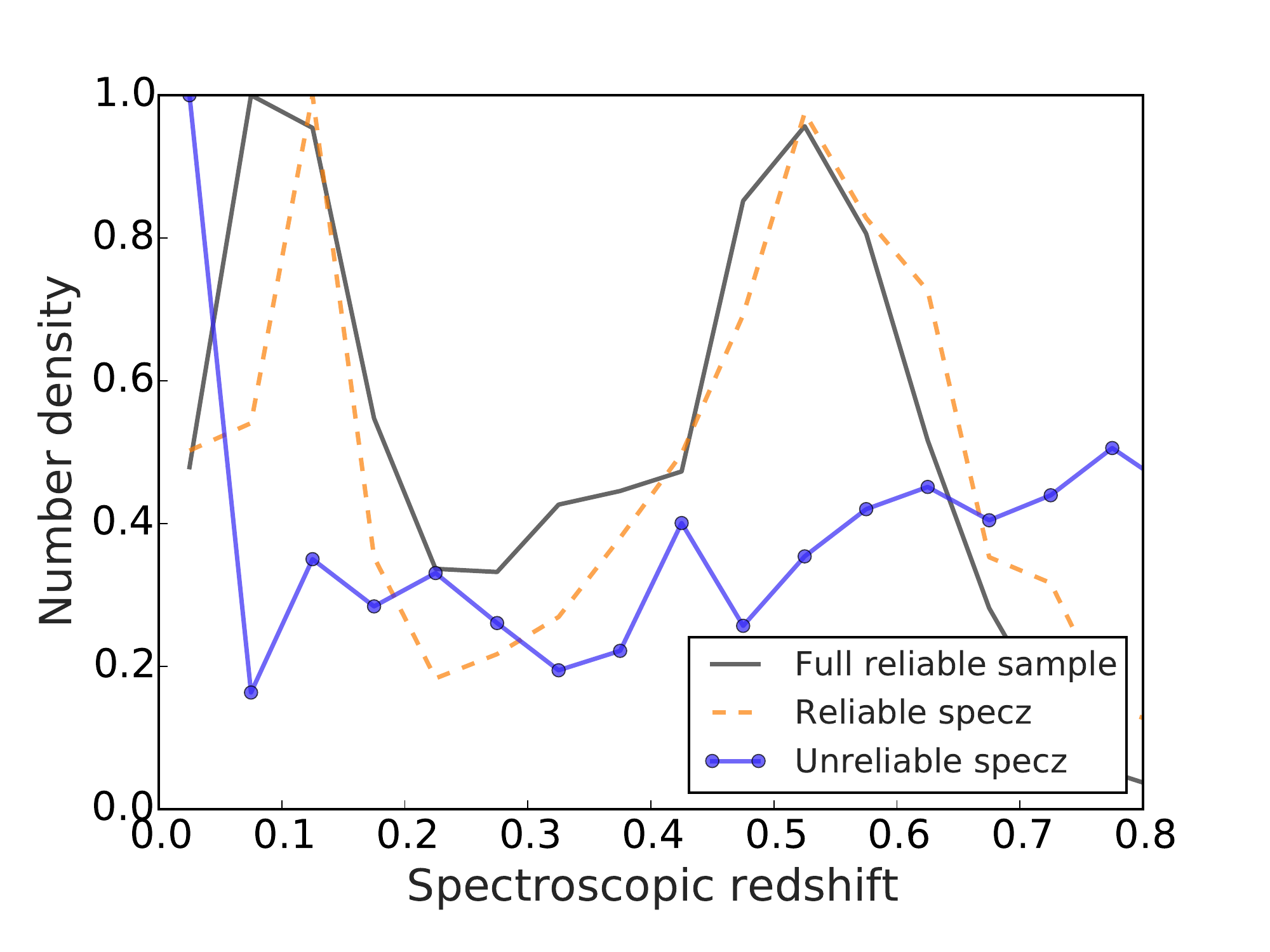}
  \caption{ \label{ABSmag} Top panel: The distribution of absolute magnitude against redshift estimated for the galaxies with both an unreliable and a reliable redshift. The reliable and unreliable data are shown by the circle and starred data points respectively. Bottom panel: The redshift distribution of the full galaxy sample by the solid grey line. The dashed orange line shows the reliable redshift distribution of those galaxies which have both a reliable and an unreliable redshift. We will remove these galaxies from the base sample. The joined dotted blue line shows the distribution of unreliable redshifts for those galaxies which we have just removed. We use these galaxies with their unreliable redshift estimates to contaminate the base sample.
}
\end{figure}

The SDSS is well suited to the analyses presented in this paper due to the enormous number of photometrically selected galaxies with spectroscopic redshifts to use as training and test samples. In particular if a galaxy spectra is obtained and subsequently found to be erroneous by the processing pipeline, the  flag `zWarning' is set to be larger than 0. For the SDSS dataset the quality flag zWarning is a good estimator of the reliability of the spectroscopic redshift. This is not always true for other spectroscopic surveys or datasets e.g. PRIMUS \citep[e.g. see Bonnet et al in prep.][]{2011ApJ...741....8C,2013ApJ...767..118C}. Furthermore the SDSS galaxies which have unreliably measured redshifts are often followed up at a later date and new spectra are obtained. Many of these new spectra often do not incur warnings during processing. It is exactly these cases which are utilized in this paper. Firstly we identify galaxies with at least one poorly measured spectrum and one well measured spectrum. Then we extract all occurrences of these galaxies from the base sample. We then assign the unreliably measured spectroscopic redshift to the galaxy and then contaminate the clean base sample with this galaxy. Finally we use machine learning to try to identify the galaxies which have unreliable spectroscopic redshifts  from those with reliable redshifts.

We select all objects from CasJobs with both spectroscopic redshifts and photometric properties which are classified as galaxies by the photometric pipeline. This sample will also include some contamination from stars and quasars. In detail we run the MySQL query shown in the appendix. The query extracts 2.5M galaxies with a range of photometric and spectroscopic qualities. The data selection is very relaxed in terms of allowed measured errors in both photometric and spectroscopic properties. In \S\ref{cleanGals} we perform a similar analysis to that which follows but impose a more stringent set of selection criteria. This query also obtains galaxies with multiple spectra measurements and allows us to identify 76639 unique galaxies with `zWarning'$>0$. Of these galaxies 9115 galaxies have both a poorly measured spectroscopic redshift above 0, and a well measured spectroscopic redshift with an error less than 0.001. We next select galaxies which have a difference in poorly measured and well measured redshifts to be greater than $0.01$ resulting in 6734 galaxies of which 3502 are unique. We impose this selection because we do not expect the error on the machine learning redshift estimate to be below 0.01. 

We use the SDSS k-correct \citep[][]{2007AJ....133..734B} package to estimate the absolute R band magnitude of the 6734 galaxies assuming both the reliable and unreliable spectroscopic redshifts. We present the distribution of absolute R magnitude against redshift in the top panel of Fig. \ref{ABSmag}, and mark the reliable and unreliable data by the circle and starred data points respectively. The bottom panel of Fig. \ref{ABSmag} shows the redshift distribution of the full galaxy sample by the solid grey line. The dashed orange line shows the reliable redshift distribution of those galaxies which have both a reliable and an unreliable redshift. We will remove these galaxies from the base sample. The joined dotted blue line shows the distribution of unreliable redshifts for those galaxies which we have just removed. We will use these galaxies with unreliable redshifts later to contaminate the base sample.

The top panel of Fig. \ref{ABSmag} shows that both the redshift distribution and the absolute magnitude distribution of the galaxies with reliable and unreliable redshift estimates are very different. The unreliable spectroscopic redshift distribution is peaked at higher redshifts. The distribution of unreliable data is also peaked at brighter absolute magnitudes. This is because the photometrically measured apparent magnitudes are unchanged, and therefore the offset is correlated with redshift. The bottom panel shows that the sample of galaxies with both reliable and unreliable redshifts are representative  of the full base sample. As expected the unreliable redshift distribution appears to be very different from the redshift distribution of the base sample.

In the analysis which follows we construct two training samples. The first is drawn from the base sample of data with reliable redshifts which has then been contaminated with anomalous data that has unreliable redshift estimates. The second training system is the first sample with a preprocessing step to remove anomalous data. We describe the method to pre-process the data in \S\ref{anomaly}. Finally we construct a validation, or `test' sample which is not used during training. The validation sample is always drawn from a non-overlapping set of base data which have reliable redshifts. We describe the construction of these samples in more detail in \S\ref{anomaly1}.

In this work we have concentrated on the following eight features for outlier estimation; the spectroscopic redshift and error, $r$ band magnitude, the following colors: $g-i$,$g-r$,$r-i$,$z-r$, and the Petrosian radius measured in the $r$ band. Of course will only use the photometric quantities when estimating redshifts. Previous work has shown that there are many other readily obtained photometric features which also have strong predictive power when estimating redshifts \citep[][]{2015MNRAS.449.1275H}. 

\subsection{Anomaly identification}
\label{anomaly}
We use the robust scikit-learn \citep[][]{scikit-learn} package called Elliptical Envelope  routine as our base anomaly detector. We briefly describe the routine below, and refer the reader to \cite{WICS:WICS61} for a review.

The Elliptical Envelope routine models the data as a high dimensional Gaussian distribution with possible covariances between feature dimensions. In short it attempts to find an boundary ellipse that contains most of the data. Any data outside of the ellipse is classified as anomalous.  The Elliptical Envelope routine uses the FAST-Minimum Covariance Determinate \citep[][]{Rousseeuw:1999:FAM:331435.331458} to estimate the size and shape of the ellipse.

In detail, the FAST-Minimum Covariance Determinate routine selects non overlapping subsamples of data and computes the mean $\vec{\mu}$, and covariance matrix $C$, in each feature dimension for each subsample. The Mahalanobis distance d$_{MH}$, is computed for each multidimensional data vector $\vec{x}$, in each subsample and the data are ordered ascendingly by d$_{MH}$. The Mahalanobis distance is defined by
\begin{equation}
\textrm{d}_{MH} = \sqrt{ (\vec{x}-\vec{\mu})^\textrm{T} C^{-1} (\vec{x}-\vec{\mu}) }
\label{eq:RRI}
\end{equation}
which reduces to the Euclidean distance if the covariance matrix is the identity matrix, and the normalised Euclidean distance if the covariance matrix is diagonal. To summarise; the Mahalanobis distance measures how many `sigma' a data point is from the mean of a distribution.

The FAST-Minimum Covariance Determinate method continues by selecting subsamples from the original samples, with small values of d$_{MH}$. The mean, covariance, and the values of  d$_{MH}$ of the subsamples are again computed. This procedure is iterated until the determinate of the covariance matrix converges.  The covariance matrix with the smallest determinate from all subsample forms an ellipse which encompasses a fraction of the original data. Data within the ellipse surface are labeled as `inliers', and data outside of the ellipse are labeled as `outliers' or anomalous, which may then be removed.

The hyper-parameter of the Elliptical Envelope routine is the contamination rate $n_c$, which is the apriori assumed fractional contamination rate of the data sample. We explore this hyper-parameter in our subsequent analysis, but note that this parameter does not need to be known to high accuracy before using the routine. We further present a method to estimate the contamination fraction from the data in \S\ref{estconfrac}. The contamination rate hyper-parameter $n_c$, describes approximately how much of the data sample should sit outside of the enclosing high dimensional ellipse that contains the majority of the data.

\subsection{Tree based methods}
One of the machine learning architectures to estimate galaxy redshifts used in this work is the scikit-learn implementation of decision trees for regression \citep[][]{ig}. The tree based machine learning architecture recursively partitions the input feature dimensions into an increasing number of bins. Each bin is chosen to minimize the scatter of the output feature, which for these purposes is the spectroscopic redshift. This results in data with very similar spectroscopic redshifts being within the same, or possibly nearby bins.

The power of tree based methods is enhanced by combining many trees. One technique to do this is called Adaptive Boosting or Adaboost \citep[][]{Freund1997119,Drucker:1997:IRU:645526.657132} which adds trees sequentially to generate an ensemble of trees. In the following we will refer to this technique as simply `Adaboost'. Adaboost weighs each new tree by its ability to predict redshifts correctly, and decides how new  trees are grown such that redshift estimates are improved for the data with poorly estimated redshifts.  For more details about combining trees with Adaboost we refer the reader to \cite{hastie01statisticallearning}\footnote{\url{statweb.stanford.edu/~tibs/ElemStatLearn}}.

In this work we choose to fix the hyper-parameter set for a single decision tree and the final number of trees and the method of growing trees. We choose the number of data on each leaf node to be 10, and the number of trees to be 100. For Adaboost we select the linear loss function, but we find that using the exponential loss function does not change the results significantly. We choose the linear loss function because the exponential loss function has previously been shown to be sensitive to label noise in classification problems \cite[][]{Dietterich:2000:ECT:350128.350131}. We note that the best machine learning hyper-parameters are normally tuned by using a cross validation sample. We note that tuning the hyper-parameters of the model can have a large effect on the machine learning redshift predictions.

\subsubsection{Mean and median regression}
We also explore a type of tree based machine learning architecture called Quantile regression, which can include the use of the median value, as opposed to the mean value when constructing the loss function for regression trees.  We use the scikit-learn package called GradientBoostingRegressor \citep[][]{Friedman99stochasticgradient,friedman2001} which accepts a parameter to determine type of loss function, for example `least squares' corresponding to mean regression, and `quantile' with a corresponding value of 50\% for median regression. For both mean and median regression we again fix the hyper-parameters of the machine learning architecture to be the same as that of the section above, except for the choice of loss function.

The loss function $L(u)$, is the method that the learning algorithm uses to find the best fitting model parameters. For trees the best fitting parameters can be the numerical values along the feature dimensions at which a split is chosen. The mean regression loss function is the least squares loss function given by
\begin{equation}
L(u) = \frac{1}{N}\sum_{i=0}^{N}\big( y_i - u \big)^2  \,
\label{eq:LSQ}
\end{equation}
where the sum runs over each of the $N$ data $y_i$ on each leaf node on the tree. The least squares loss function is sensitive to outliers, and so we would expect it to be more affected by outliers in the training set. For median regression the loss function is given by
\begin{equation}
L(u)=\frac{0.5}{N}\big(-\sum_{y_{i}<u}(y_{i}-u)+\sum_{y_{i}\geq u}(y_{i}-u)\big) = \frac{1}{N}\sum_{i=0}^{N} \big| y_{i}-u \big| \,
\end{equation}
which is less sensitive to outliers because of the linear dependence on the differences between values $y_i$ and $u$. We compare the results of training these different architectures on contaminated, and outlier removed data samples in \S\ref{meanMedRegression}.

\subsection{Self Organising Maps}
Another popular machine learning architecture is the Self Organising Map \citep[][hereafter SOM]{Kohonen:1997:SM:261082}, which have recently been used for redshift estimation \citep[][]{Geach21012012}. SOMs are also being used in combination with template fitting routines for photometric redshifts (Greisel et al in prep). We use the public implementation of a SOM, called SOMz \citep[][]{2014MNRAS.438.3409C}\footnote{github.com/mgckind/MLZ} which we briefly describe below. We refer the reader to \cite{2014MNRAS.438.3409C} for more details. {\rr We choose to include SOMz in this paper because it represents a very different machine learning architecture than those of tree based methods. Using both SOMs and trees suggests how generalisable the results found in this paper are.}

SOMz combine neural networks with dimensionality reduction and similarity clustering. The SOMs are evolved from random starting weights such that training examples with similar high dimensional inputs appear clustered in a two dimensional space of pixels. The map evolution is unsupervised because it is performed by only examining the input features. Once the SOMz is stable, the training examples are again passed through the map, and the values of the output feature are combined to produce an output value for each pixel. New data are passed through the SOMz and the pixel, or nearby pixels, which have the largest activation values contribute to the predicted value returned.

In this work we choose to fix the hyper-parameters of the SOMz to have a spherical map geometry with 768 ($=12\times8\times8$) pixels and we perform 100 training iterations. For a full analysis on the effect of using different hyper-parameters with SOMz see \cite{2014MNRAS.438.3409C}. Again we mention that tuning these hyper-parameters can lead to a large amount of improvement, however this is not the focus of this work.

\section{Analysis and Results}
\label{results}
We first introduce the anomaly detection method to both identify the inserted contaminating galaxies with unreliable redshifts, and then to build a cleaner training sample. We then provide a method to estimate the contamination fraction of a dataset. Finally we train separately on both the full contaminated sample, and the cleaned sample, and show the effect on the measured statistics of the machine learning redshift as calculated on an independent and single cross validation sample.

\subsection{Anomaly identification}
\label{anomaly1}
We examine the ability of the Elliptical Envelope method to correctly identify the galaxies with unreliable redshift estimates that we use to contaminate the training sample. We perform more than 250 sets of independent analysis for both the Adaboost, and SOMz machine learning architectures.  We initial randomly select a number $N_{ur}$, where 100$<N_{ur}<$6730, of galaxies with unreliable redshifts, and combine them with $N_r$ randomly selected galaxies with reliable redshifts from the base sample. We restrict $N_r$ to values $3*N_{ur}<N_r<$100k. 

We then construct a training and cross validation sample from this combined sample and perform feature normalization on all of the features. Throughout this paper we ensure that the cross validation sample is only drawn from those galaxies with a reliable redshift estimate, because it would be irrelevant to try to predict the redshift of galaxies with unreliable redshift estimates.  For this training sample we explore a range of values of the hyper-parameter $n_c$, ranging from $10^{-5}<n_c<0.5$, corresponding to different initial `best guesses' of the expected contamination fraction as used by the Elliptical Envelope routine.

For each value of $n_c$ the anomaly detection code produces a classification of either `inlier' or `outlier' for each galaxy. We determine the percentage of correctly identified outlier galaxies which have an unreliable redshift, and also the percentage of potentially incorrectly identified galaxies with a reliable redshift. We note that the galaxy sample with reliable redshifts may however be an outlier along a different feature dimension other than the spectroscopic redshift. 

\begin{figure}
 \includegraphics[scale=0.47,clip=true,trim=0 15 50 30]{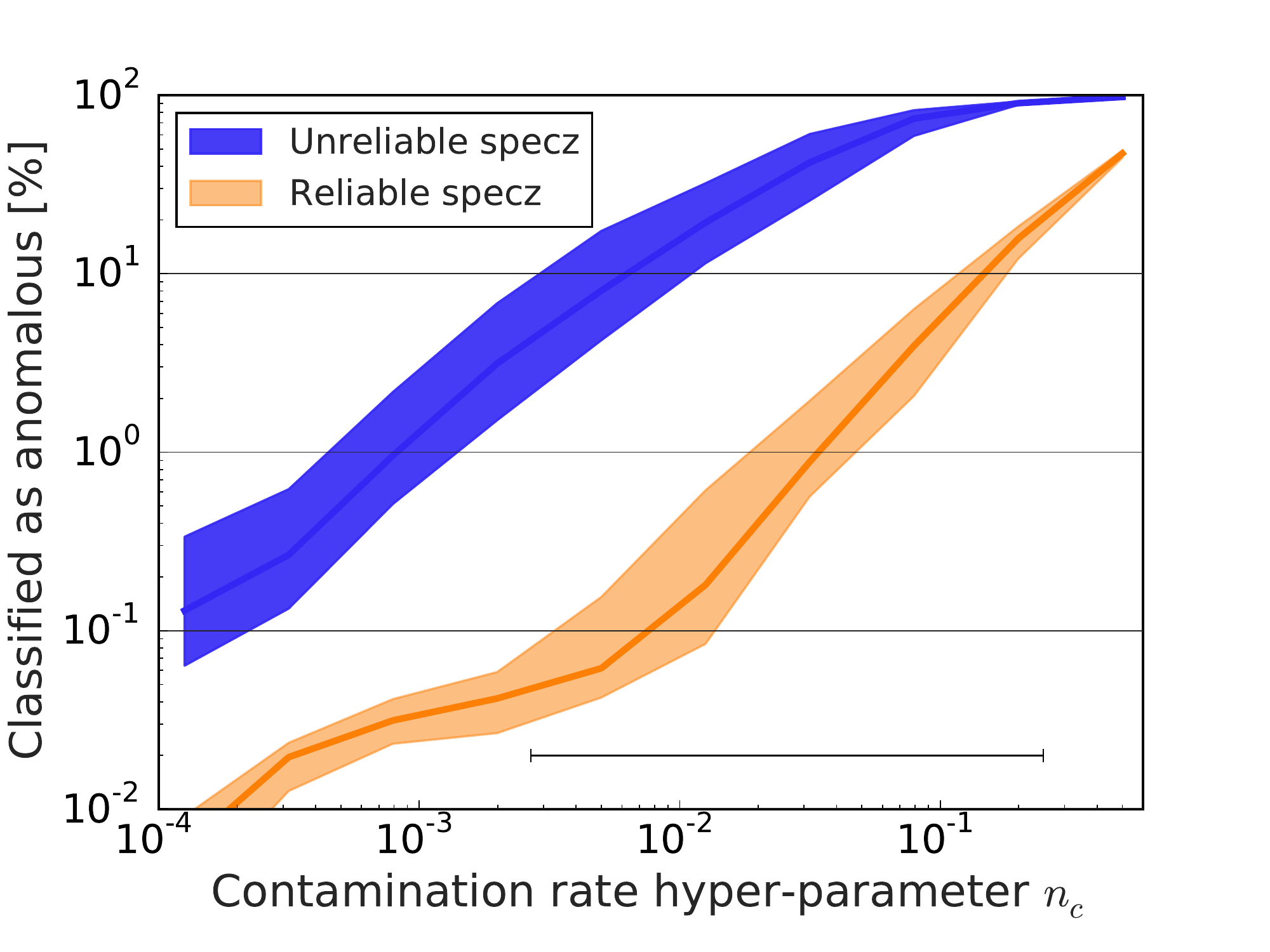}
  \caption{ \label{idError} The percentage of correctly identified outlier galaxies as a function of the  hyper-parameter $n_c$, which measures the input contamination fraction best guess. The dispersion of data at fixed $n_c$ is due to the different randomly selected combined samples of random size. The dark lines show the mean of the distribution and the upper and lower shaded regions show the 68\% spread of the data. The black error bar show the actual range of contamination fractions, which correspond to the number of galaxies with unreliable redshifts inserted into the base sample. Each of the 250 experiments has a different inserted contamination fraction.
}
\end{figure}

In Fig. \ref{idError} we show the percentage of correctly identified galaxies with unreliable redshifts  as a function of the contamination hyper-parameter $n_c$. The dispersion of data at fixed $n_c$ is due to the different randomly selected combined samples of random size. The dark lines show the mean of the distribution and the upper and lower shaded regions show the 68\% spread of the data. The black error bar shows the actual range of contamination fractions, that correspond to the number of galaxies with unreliable redshifts inserted into the base sample. Each of the 250 experiments has a different inserted contamination fraction.

We find that the fraction of data with unreliable redshifts which is classified as anomalous, or an outlier, is between one and two orders of magnitude larger than the corresponding fraction of data with reliable redshifts. This demonstrates the success of the Elliptical Envelope technique to identify data with unreliable redshift estimates.

\begin{figure}
   \centering
   \includegraphics[scale=0.47,clip=true,trim=0 15 40 30]{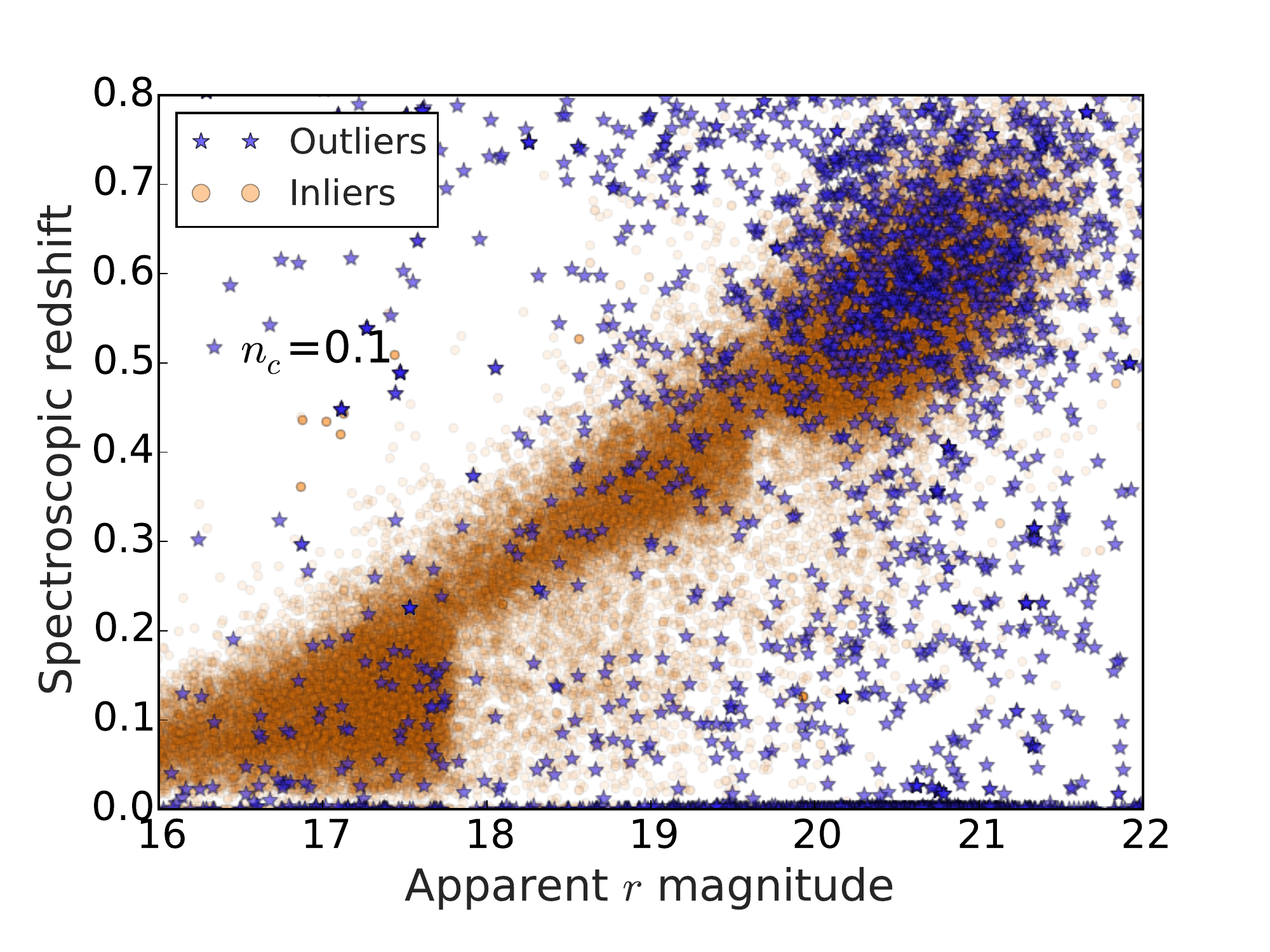}
   \includegraphics[scale=0.47,clip=true,trim=0 15 40 30]{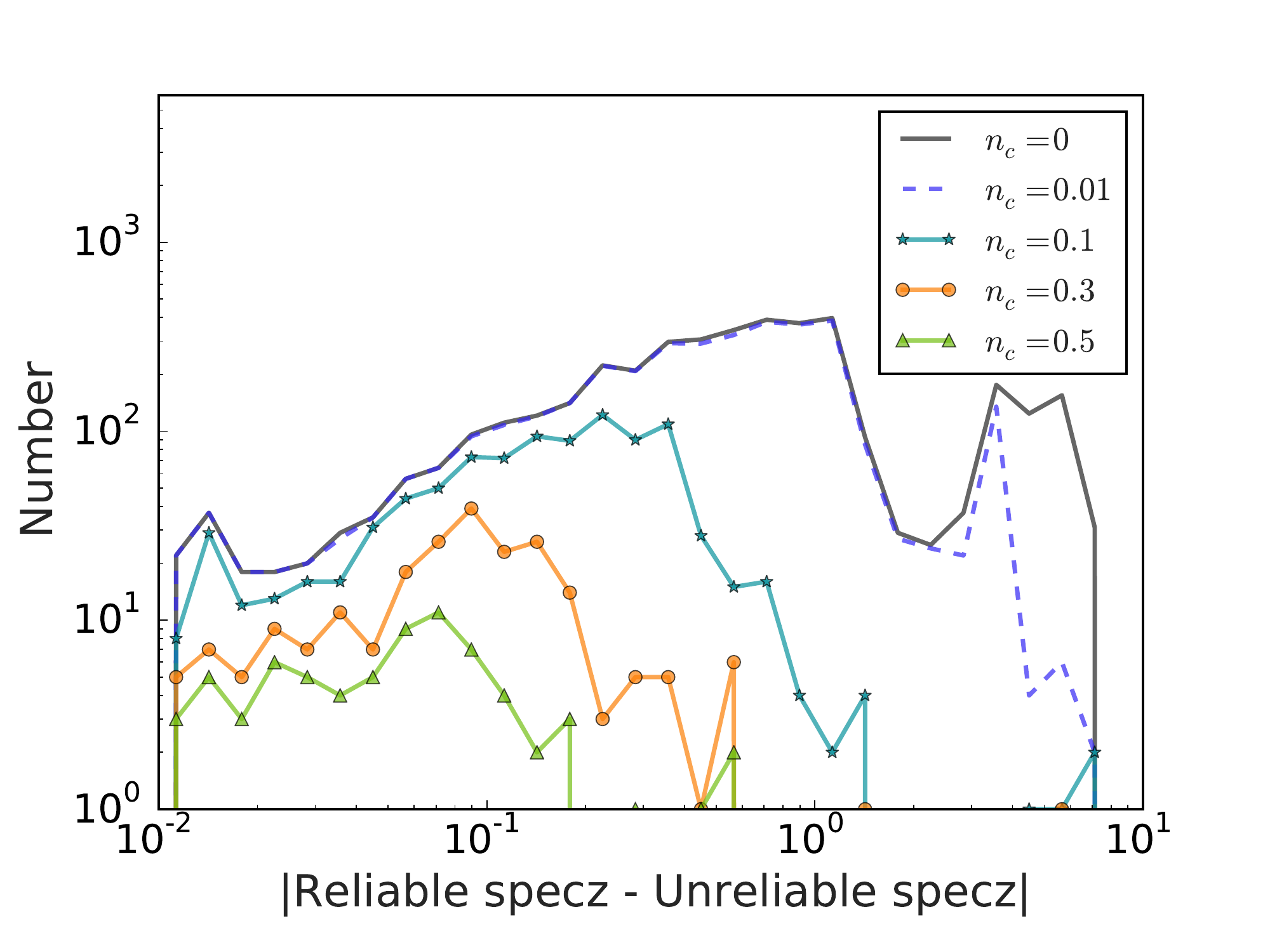}
   \caption{ \label{distOutliers}
   Top panel: The transparent circles show the redshift and apparent magnitude distribution of the base sample contaminated with unreliable redshifts. The blue stars show which of those galaxies are classified as being outliers using the Elliptical Envelope technique for a given contamination hyper-parameter value $n_c=0.1$. Bottom panel: The absolute difference between the reliable and unreliable redshifts for the contaminating galaxies which are {\it not} classified as being outliers by the Elliptical Envelope technique as a function of increasing $n_c$.} 
\end{figure}

We next explore which of the contaminating data with unreliable redshifts is classified as anomalous, and show projections through the data in Fig. \ref{distOutliers}. In the top panel of Fig. \ref{distOutliers}, the transparent circles show the redshift and apparent magnitude distribution of the base sample contaminated with unreliable redshifts. The blue stars show which of those galaxies are classified as being outliers using the Elliptical Envelope technique for a given contamination hyper-parameter value $n_c=0.1$. The bottom panel of Fig. \ref{distOutliers} concentrates on those contaminating galaxies with both a reliable and unreliable redshift. The panel shows the absolute difference between the reliable and unreliable redshifts for the contaminating galaxies which are {\it not} classified as outliers by the Elliptical Envelope technique as a function of increasing $n_c$.

The top panel of Fig. \ref{distOutliers} shows that galaxies which occupy a region of redshift and $r$ band apparent magnitude space which is very different from the majority of other galaxies are classified as being anomalous. We also note that a small fraction of galaxies which occupy the same region of redshift and $r$ band apparent magnitude space as the majority of galaxies, is also classified as anomalous. This could be because the data is anomalous along one or more different feature dimensions, which is not easily viewed in this two dimensional projection. There are three distinct clouds of data with reliable redshifts in the top panel. These clouds of data correspond to the different observing phases of the SDSS.

The bottom panel of Fig. \ref{distOutliers} shows that the number of galaxies with unreliable redshifts which are {\it not} classified as anomalous decreases as the contamination fraction hyper-parameter $n_c$ increases. We also note that the most extreme examples of galaxies with very anomalous unreliable redshifts are preferentially removed as $n_c$ increases. The sharp drop at the x-axis location of 0.01 is due to the construction of the contaminating data sample.

{\rr In the top panel of Fig. \ref{distOutliers} there are distinct clouds of data in these feature projections. These are due to the different observing strategies of the SDSS. For example most of the faint, high redshift cloud were observed in SDSS III, while the lower redshift clouds were observed in SDSSI/II. We also perform outlier detection separately for these samples, and find the following similar trend in both samples: the fainter the galaxy is in the $r$ band, the more likely it is to have an anomalously large unreliable redshift. This can be understood by the fainter galaxies being more difficult to observe spectroscopically and requiring larger integration times.}

We have also explored the use of One Class Support Vector Machines \citep[][]{SVM} as the machine learning anomaly detector, but do not find an improvement over the results using the Elliptical Envelope method. This suggests that a hyper dimensional ellipse provides a good model to enclose, and therefore identify, the non-anomalous data.

\subsection{The distribution of data with anomalies removed}
\label{anomRemoved}
We explore how the distribution of galaxies changes as a function of the contamination hyper-parameter $n_c$, as compared to the initial sample. We construct a sample of size 100k which is contaminated with 3k galaxies with unreliable redshifts.

\begin{figure*}
   \centering
   \includegraphics[scale=0.32,clip=true,trim=15 15 45 30]{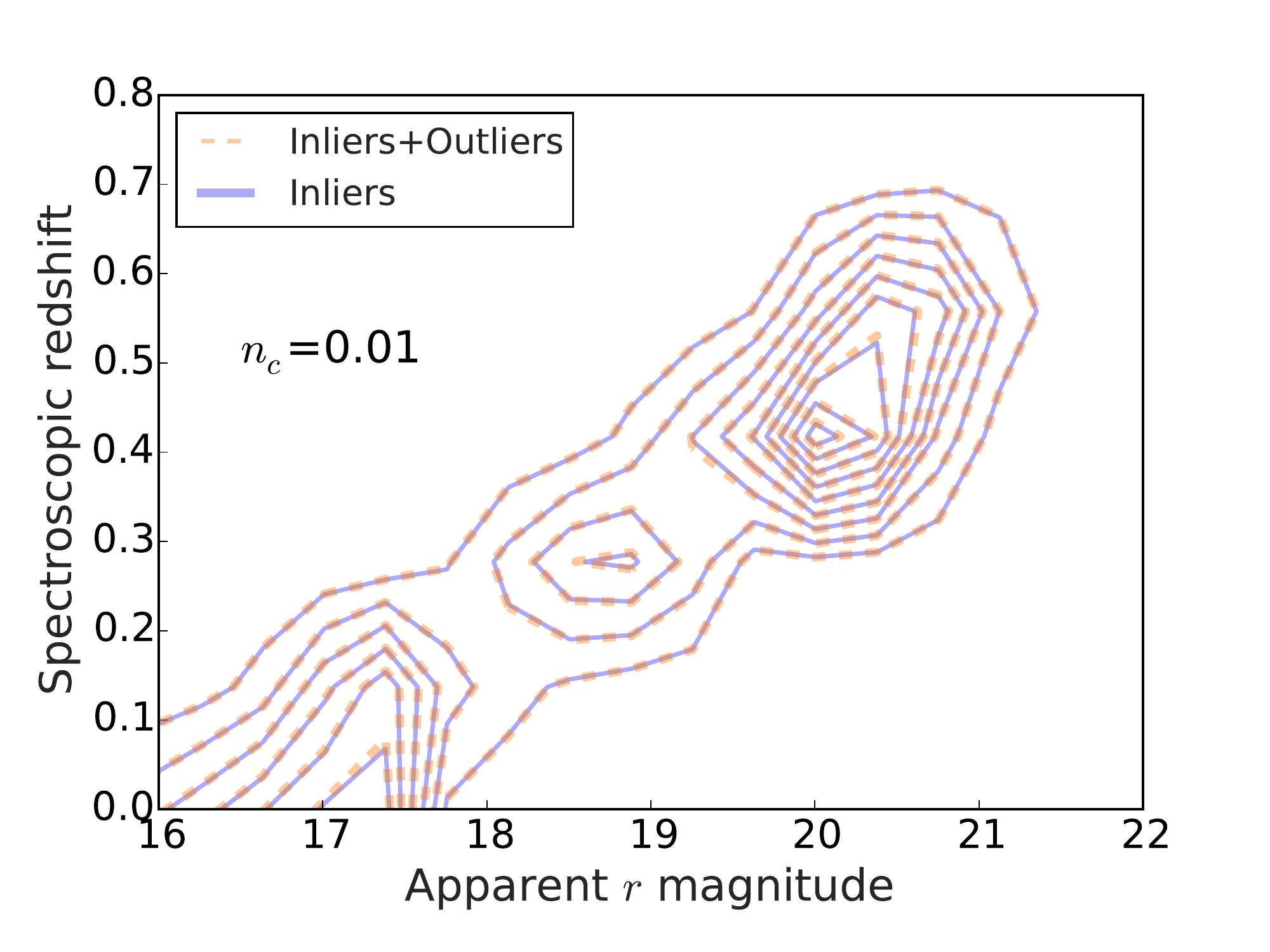}
   \includegraphics[scale=0.32,clip=true,trim=15 15 45 30]{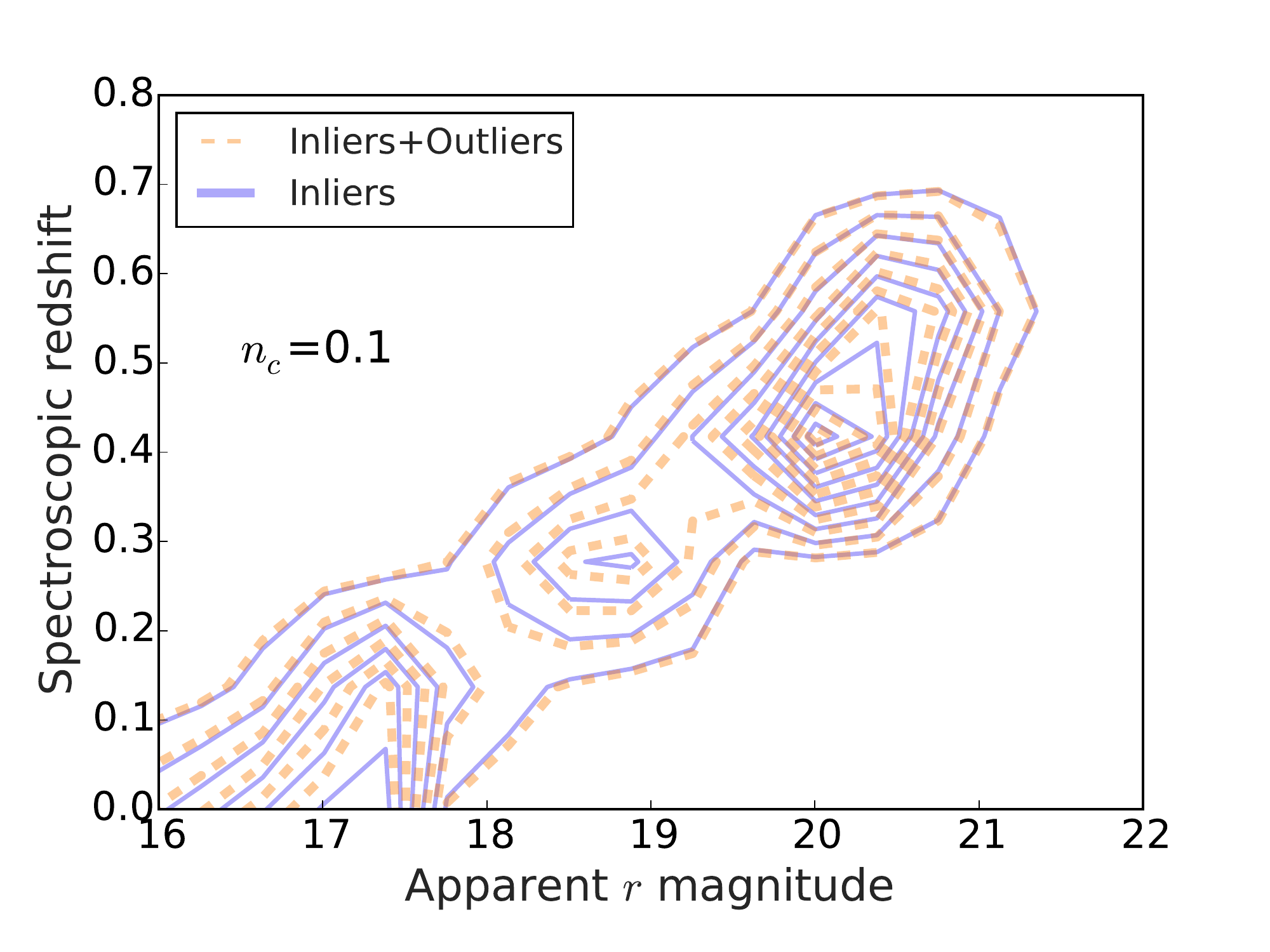}
   \includegraphics[scale=0.32,clip=true,trim=15 15 45 30]{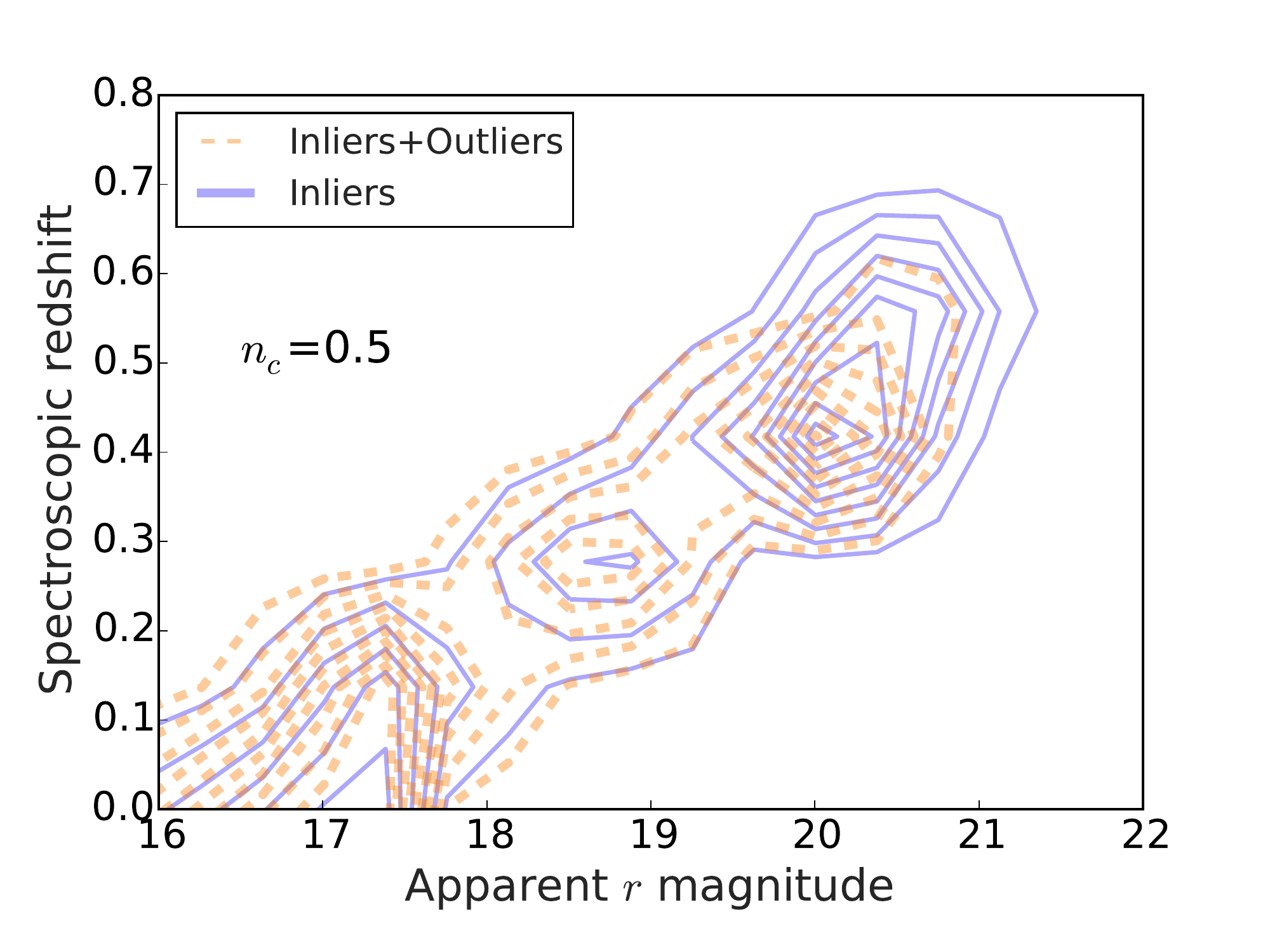}
   \caption{ \label{distClean} The distribution of training galaxies as a function of the contamination hyper-parameter $n_c$. We show the full sample by the solid lines, and the sample with `anomalous' galaxies removed by the dashed line. Each panel shows the change in the distributions when using a data sample of size 100k which has been contaminated with 3k galaxies with unreliable redshifts. } 
\end{figure*}

We perform anomaly detection on the contaminated sample for different values of $n_c$. In Fig. \ref{distClean} we show the distribution of spectroscopic redshift against apparent magnitude in the $r$ band, for three different values of $n_c$ indicated in each panel. The combined sample in each case is shown by the solid lines, and the sample with anomalous outliers removed is shown by the thick dotted lines.

Fig. \ref{distClean} shows that as the contamination hyper-parameter increases above $n\geq0.01$ so the distribution of galaxies becomes biased with respect to each other. For small values of $n_c$ the distributions are mostly unaffected. If there is no anomalous data, and the Elliptical Envelope routine is expecting a large fraction of contaminated data, then even clean data is removed, however if anomalous data is indeed present, then the routine will detect it. This behavior can also be seen in Fig. \ref{idError}.

In the next section we derive a prescription to estimate the contamination fraction from a base data sample that may be contaminated.

\subsection{Estimating the contamination fraction}
\label{estconfrac}
We next provide a prescription to make an empirical initial estimate for the contamination fraction. We note that the Elliptical Envelope method is not very sensitive to the exact value of the contamination fraction, as shown in Fig \ref{idError}, and therefore we are  interested in obtaining an order of magnitude estimate. We use the measured values of Mahalanobis distance d$_{MH}$, to estimate the contamination rate. 
 
To make this analysis more realistic we construct a base, and contaminated sample, with more stringent selection criteria on the allowed photometric and spectroscopic errors.  We select galaxies which pass the following selection criteria: measured errors in $r,g,i$ bands between $0<$error$<0.2$ and spectroscopic redshifts greater than 0 and spectroscopic redshift errors between $0<$error$<0.001$. This reduces the base sample with reliable redshifts to 2.1M and the sample with unreliable redshifts to 3017. 

For this analysis we construct 250 datasets, which again contain a random amount of data with unreliable redshifts, and a random sample of base data with reliable redshifts. We use the Elliptical Envelope technique with a range of contamination fractions $0.001<n<0.5$, to measure d$_{MH}$ of the data for each value of $n_c$. We note that the dimensionality of the input feature space N$_{D}$, is 8, as described in \S\ref{obs_data}. We then assign the class `outlier' to data that satisfies $\textrm{d}_{MH}>\textrm{N}_{sig}$. We find that the choice of $\textrm{N}_{sig} = 2^{\textrm{N}_{D}}$ provides a good estimation for the outlier fraction, and discuss the robustness of this value below. 

\begin{figure}
   \centering
   \includegraphics[scale=0.47,clip=true,trim=20 5 40 25]{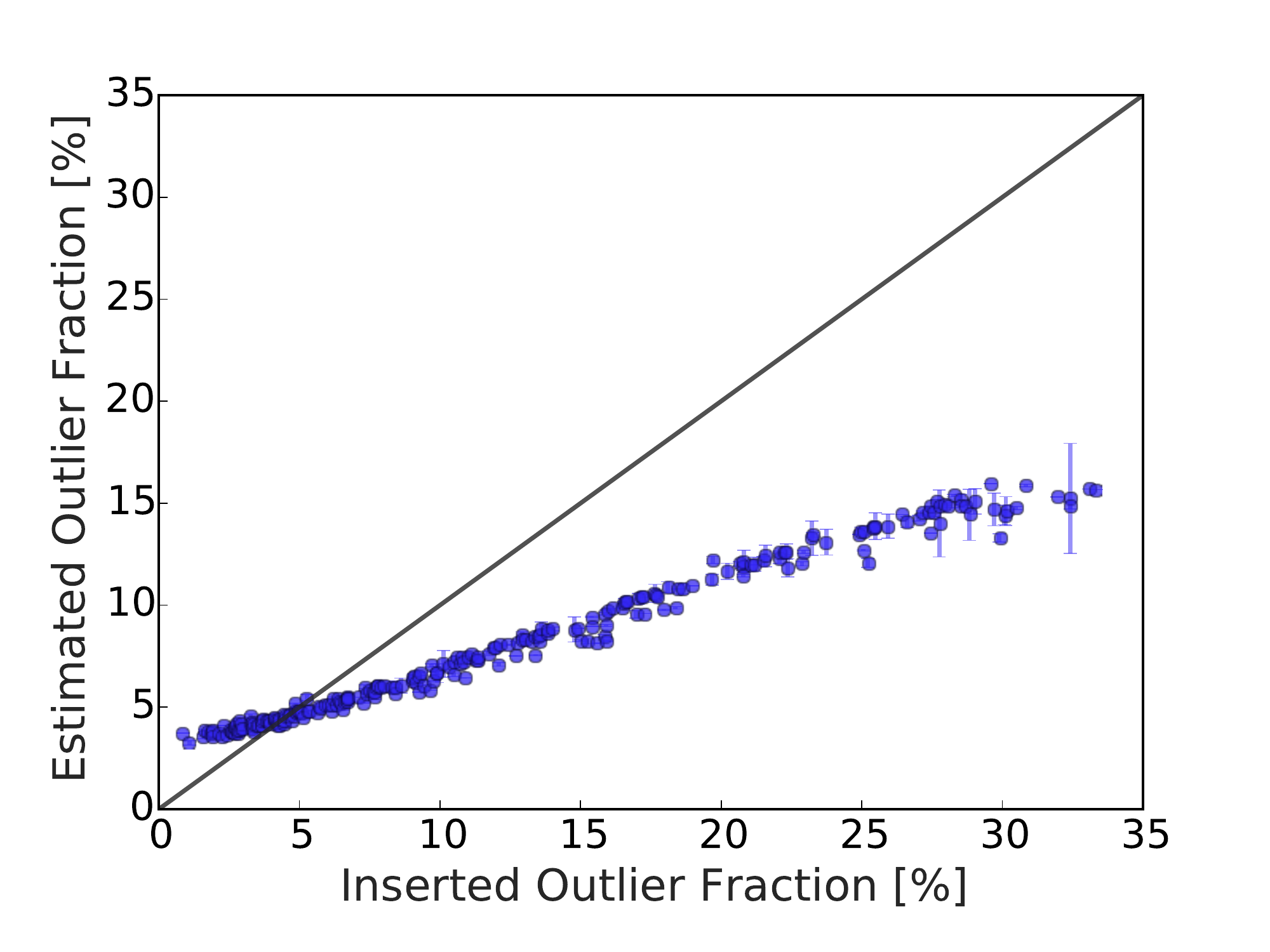}
   \caption{ \label{estContaminationFraction} The fractional contamination rate of data with unreliable redshifts inserted into the base sample, against the estimated contamination fraction using the Mahalanobis distance d$_{MH}$. We define contaminated galaxies as those that satisfy $\textrm{d}_{MH}>2^{\textrm{N}_{D}}$ for the $\textrm{N}_{D}$ feature dimensions of the data sample. The error bars are inflated by a factor of 10, and show the 68\% spread of results using different values of the contamination hyper-parameter $n_c$, when using the Elliptical Envelope technique to measure d$_{MH}$.} 
\end{figure}

Fig. \ref{estContaminationFraction} shows the fractional contamination rate of data with unreliable redshifts inserted into the base sample against the estimated contamination fraction using the Mahalanobis distance d$_{MH}$. The error bars are inflated by a factor of 10, and show the 68\% spread of results using different values of the contamination hyper-parameter $n_c$, when using the Elliptical Envelope technique to measure d$_{MH}$.

We note that a large range $1.75^{\textrm{N}_{D}}<\textrm{N}_{sig}<3^{\textrm{N}_{D}}$ of values corresponding to $88<\textrm{N}_{sig}<6560$ also produce reasonable `order of magnitude' estimates of the inserted contamination fraction. As an illustrative example we could compare this result to the case of a two dimensional Gaussian distribution of width $\sigma$; this relationship is equivalent to assigning the classification of outlier to data that is more than $4\sigma$ away from the mean value. 

\subsection{Machine learning redshifts from anomaly removed training data}
We next present the effect on the machine learning redshift if we train only on the training sample with anomalous data removed, instead of training on the full contaminated sample. We remove anomalous data using the Elliptical Envelope technique. We choose to use Adaboost and SOMz in independent sets of analyses. 

In each set of analyses we first train on the contaminated training sample, and then use the Elliptical Envelope method with a fixed contamination fraction hyper-parameter $n_c$, to remove anomalous data, irrespective of whether or not they are drawn from the sample with reliable or unreliable redshift estimates. This produces a cleaned training set, which we then independently train on. We refer to this as the `cleaned' training sample in what follows. 

We construct a cross-validation sample drawn from galaxies with reliable spectra. To make a fair comparison later, we do not modify the cross-validation sample at all, irrespective of their inlier or outlier definitions.  We then pass the same cross-validation sample through both learned systems, and obtain a machine learning redshift estimate $z$, for each galaxy. 

We construct the redshift scaled residual vector $\Delta_{z'}=$($z-$spec$z$)/(1+spec$z$) and measure the following metrics: $|\mu|,\sigma_{68},\sigma_{95}$, corresponding to the median value of $\Delta_{z'}$, and the values corresponding to the 68\% and 95\% spread of $\Delta_{z'}$, and we additionally measure the `outlier rate' defined as the fraction of galaxies for which $|\Delta_{z'}|>0.15$. Note that the outlier rate here has a different, albeit related, definition from the anomaly detection sections.  We repeat this analysis for Adaboost and SOMz, and then repeat the entire analysis for a different value of $n_c$.  We perform 250 sets of experiments, each with a randomly selected initial training sample of data with reliable and unreliable redshifts, and with a randomly selected cross-validation sample.

\begin{figure*}
   \centering
   \includegraphics[scale=0.47,clip=true,trim=0 15 50 30]{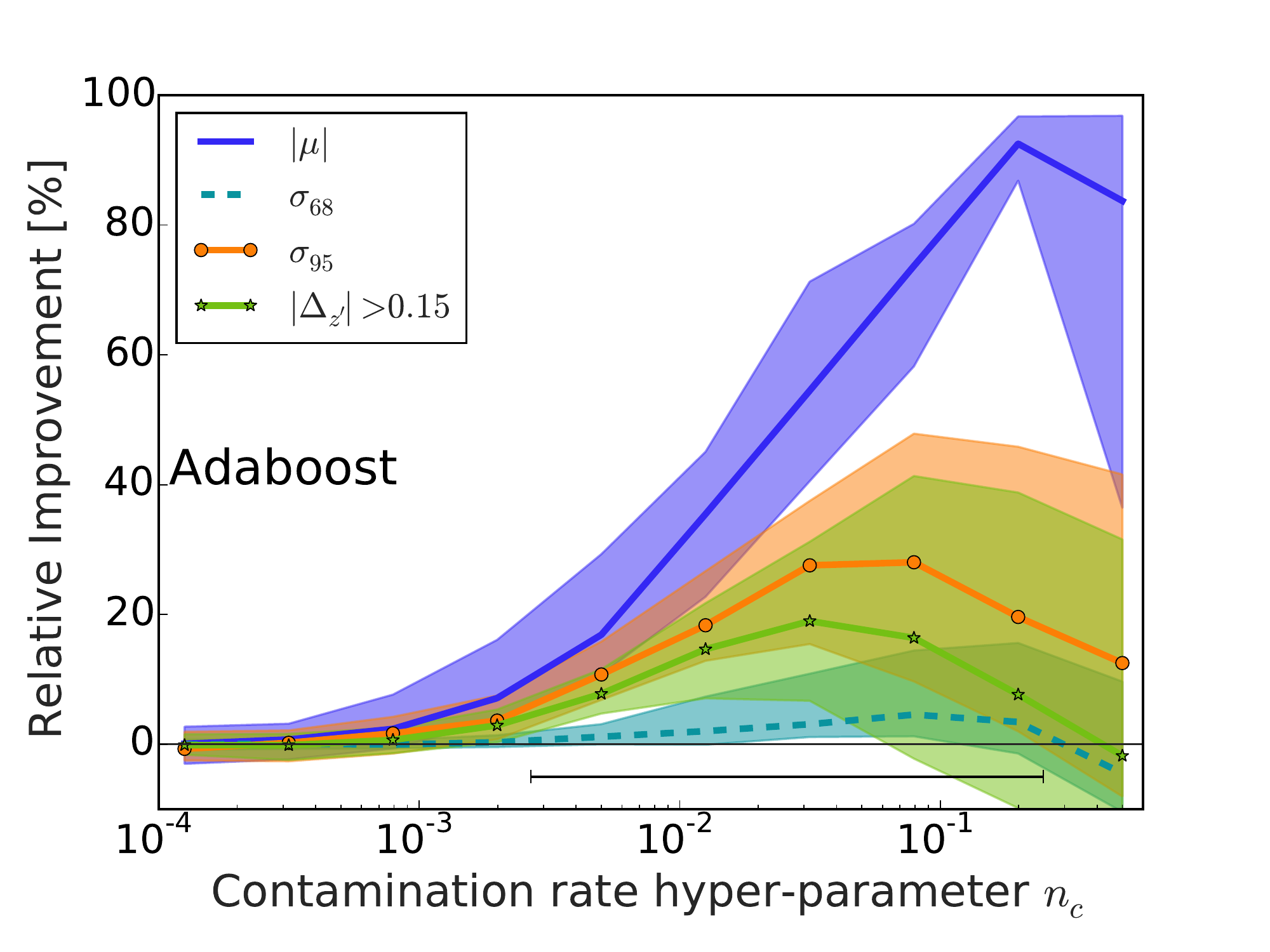}
   \includegraphics[scale=0.47,clip=true,trim=0 15 50 30]{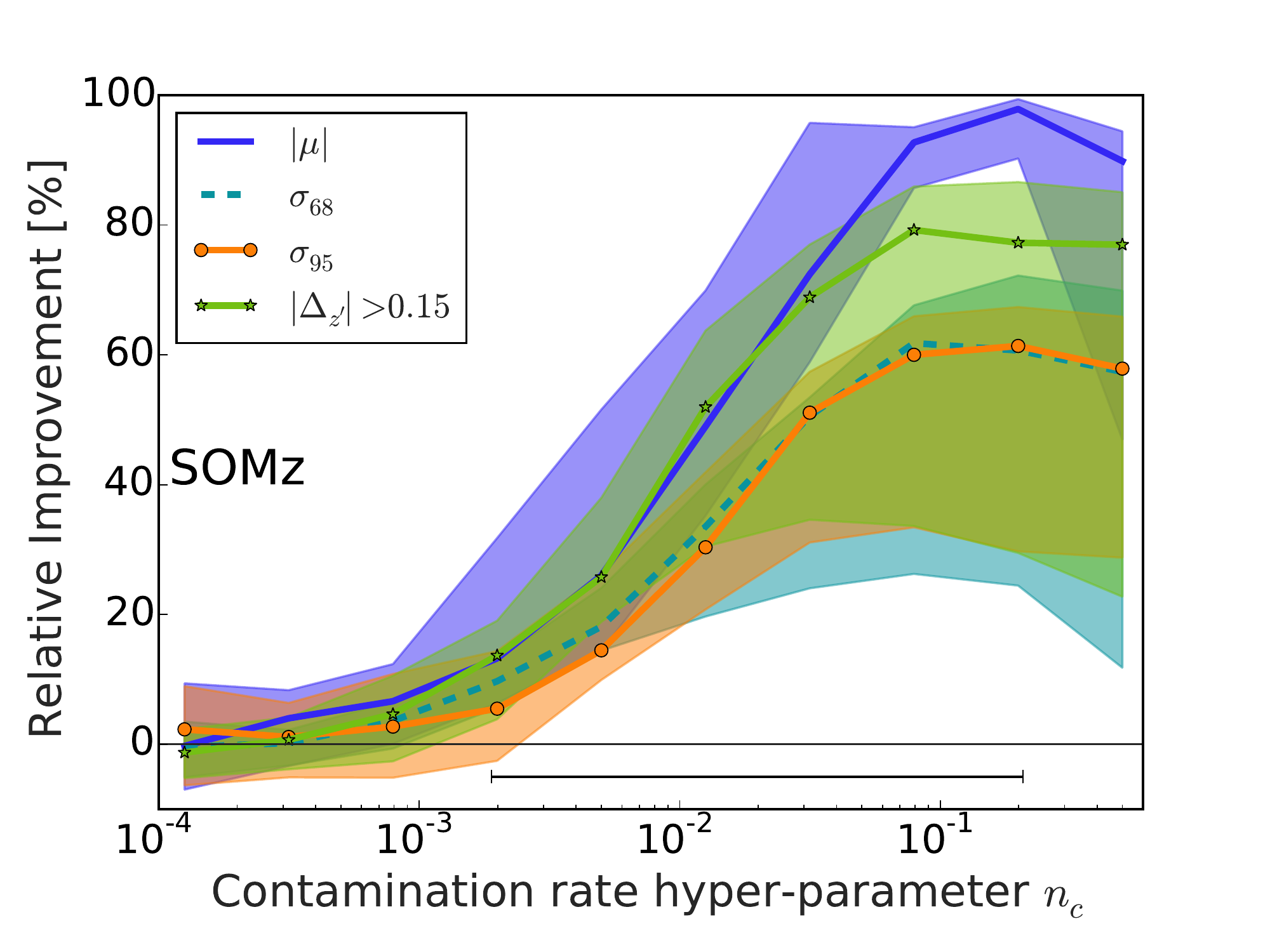}
   \caption{ \label{relativeImprovement} The percentage relative improvement when training on the cleaned sample instead of the full sample on each of the measured statistics, as a function of the hyper-parameter $n_c$. In the left hand panel we show the analysis with Adaboost, and in the right hand panel we show the improvement with SOMz. The black error bar show the actual range of contamination fractions, which correspond to the number of galaxies with unreliable redshifts inserted into the base sample. Each of the 250 experiments has a different inserted contamination fraction.} 
\end{figure*}

In Fig. \ref{relativeImprovement} we show the percentage relative improvement when training on the anomaly cleaned sample instead of the initial contaminated sample on each of the measured statistics, as a function of the hyper-parameter $n_c$. In the left hand panel we show the results of the analysis with Adaboost, and in the right hand panel we show the results with SOMz. The lines and shaded regions again corresponds to the median and 68\% of the distribution.

In both sets of analysis we find that for very small values of $n_c<0.001$, corresponding to a removal of 1\% of data with unreliable redshifts, and 0.05\% of data with reliable redshifts (see Fig. \ref{idError}), we find a small improvement in the measured metrics at the level of a few percent or less. For increasing values of $n_c$ to 0.07, corresponding to a removal of 70\% of unreliable data and 3\% of reliable data, we find the improvement in the metrics for both machine learning systems with values between 20\% and 80\%. The metrics most affected by the removal of anomalous data are the median values, and the tails of the distribution, namely $\sigma_{95}$ and the outlier fraction $|\Delta_{z'}|>0.15$. {\rr Fig. \ref{relativeImprovement} shows that there is a slight to moderate decline in improvement of the metrics at larger values of $n_c$. This degradation in improvement can be understood by examining the effect of large $n_c$ on the resulting distributions of training galaxies as see in Fig. \ref{distClean}. For larger values of $n_c$ the cleaned samples become less representative of the initial sample, and therefore the training and test sets become less representative of each other, and the machine learning mapping extends into the realm of extrapolation. Extrapolating outside of the training set leads to spurious and degrading results, as seen here.}

Fig. \ref{relativeImprovement} shows the relative improvement for each of the two machine learning techniques. We also perform a comparison between these two machine learning architectures and show the results in the top two rows of Table \ref{photoRapter}. We note that this is not the main objective of this work because similar comparisons have already been performed \citep[e.g.,][]{2014MNRAS.438.3409C}. The table shows the machine learning architecture used and the effect of training on both the data sample that is contaminated  with unreliable redshifts, and the data sample with outliers removed using the Elliptical Envelope technique. We show the the measured statistics in the final four column headings. The quoted values are the median values at fixed $n_c=0.07$ of the 250 samples that each have a different inserted contamination fraction. We note that Adaboost outperforms the SOMz algorithm on all metrics by a factor of $>2$ when training on the contaminated samples, and is comparable with or outperforms the SOMz algorithm when training on the cleaned samples. We have chosen to show the  results obtained for a contamination hyper-parameter value $n_c=0.07$, but note the same behavior is found for all values of $n_c$.

We note that both panels of Fig. \ref{relativeImprovement} show improvement as the base sample is cleaned of contaminating data. This shows that the machine learning routines for which the improvement is the greatest, are the least robust techniques to use when presented with some fraction of anomalous training data. We further explore other techniques which are less susceptible to anomalous training data in \S\ref{meanMedRegression}.

During this work we assume that the cross validation sample is not contaminated with anomalous data, which is true by construction. However this may not be true of other data sets. In such cases one could perform anomaly detection on both the training, cross validation, and test sets to remove outliers from the full sample. If the sample anomaly detection results were applied to a final test sample, this would result in a fair analysis. However one would need to check that this preprocessed data is suitable for the final science application at hand. One further method would be to identify anomalous cross validation data, and then investigate these data to understand why they have been so classified.

\subsubsection{Mean vs Median regression}
\label{meanMedRegression}
\begin{figure*}
   \centering
   \includegraphics[scale=0.47,clip=true,trim=0 15 50 30]{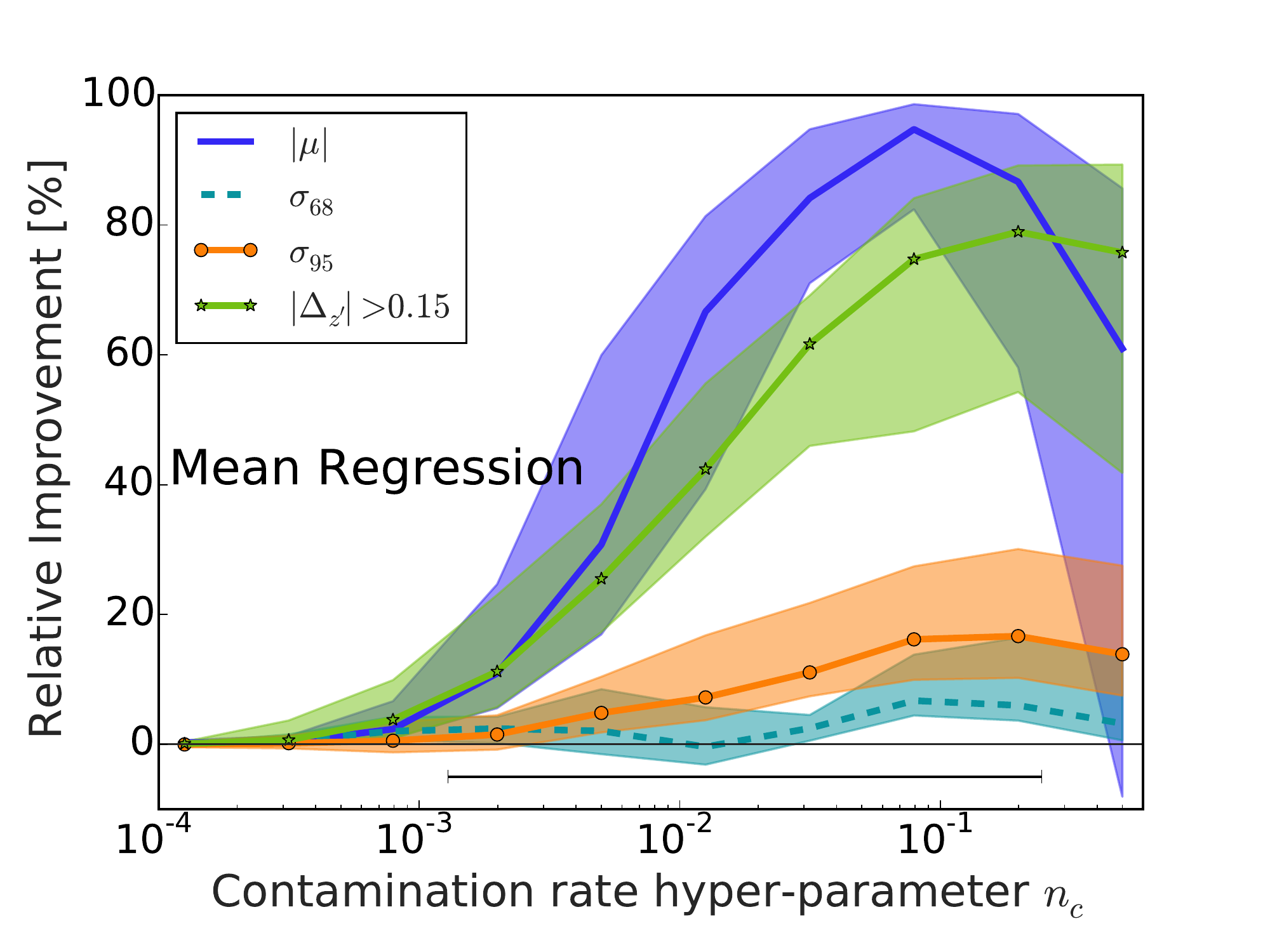}
   \includegraphics[scale=0.47,clip=true,trim=0 15 50 30]{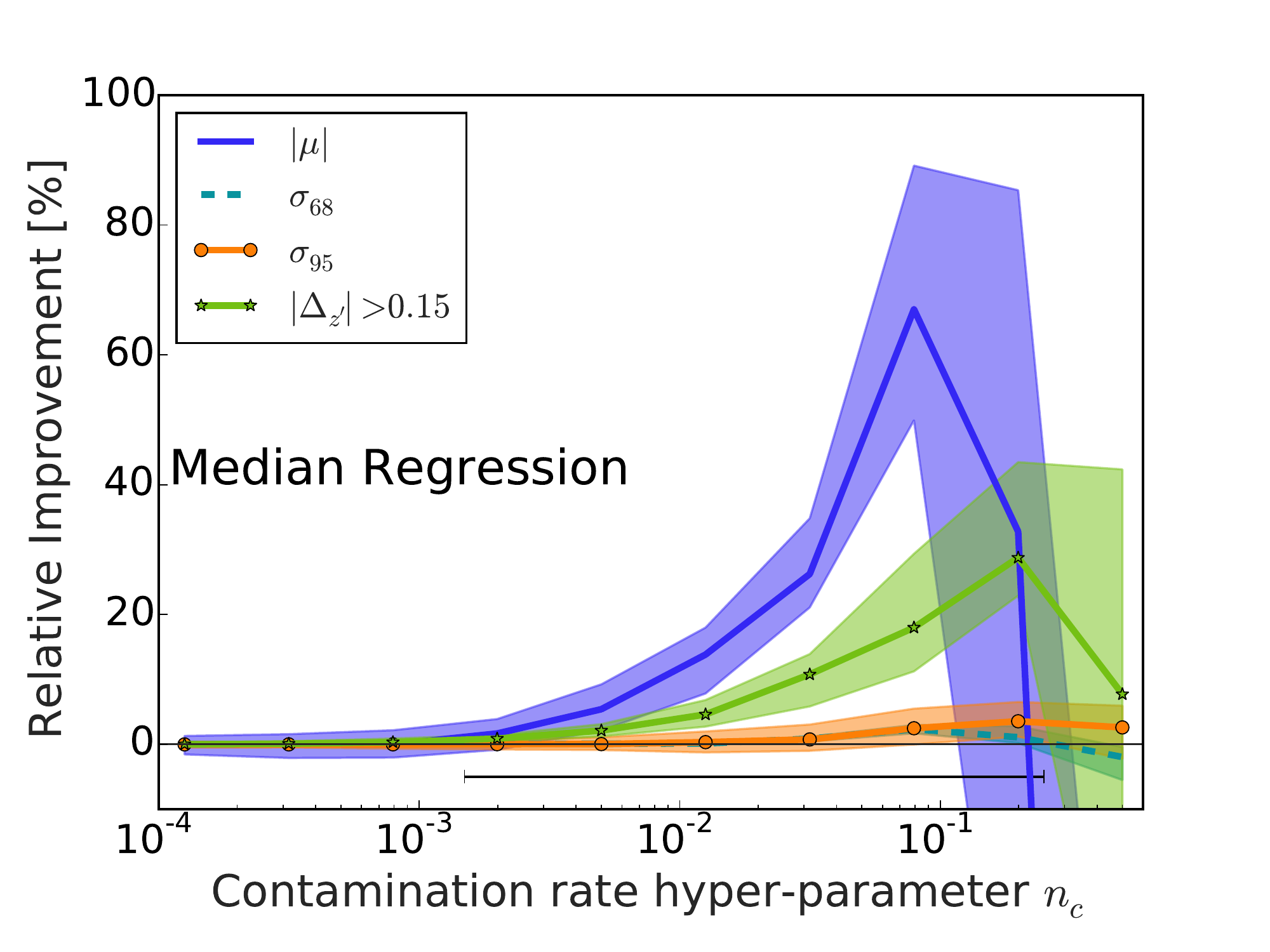}
   \caption{ \label{relativeImprovemenMeanMedian1}  These panels are similar to those in Fig. \ref{relativeImprovement} and also include a data sample contaminated with unreliable redshifts. In the left hand panel we show the analysis with mean regression, and in the right hand panel we show the relative improvement using median regression. The black error bar show the actual range of contamination fractions, which correspond to the number of galaxies with unreliable redshifts inserted into the base sample. Each of the 250 experiments has a different inserted contamination fraction.} 
\end{figure*}

We next explore the machine learning architecture called mean and Quantile, or median, regression. Quantile regression can use the median, as opposed to the mean value when constructing the loss function for boosted regression trees. We expect median regression to be less strongly affected by contamination in the training data. For comparison with \S\ref{anomRemoved} using Adaboost, we construct very similar machine learning architectures using the same hyper-parameters and only vary the loss function. We continue by applying the same formalism as before: we first train on the contaminated data sample, and then use the Elliptical Envelope method to remove outlier data, and finally retrain on the cleaned data sample. We show the results of using mean regression in the left hand panel of Fig. \ref{relativeImprovemenMeanMedian1}, and we show the results using the median regression in the right hand panel. Again we show the actual spread of the inserted contamination fraction using the galaxy sample with unreliable redshifts is shown by the black starred data point and error bar.

We find that both machine learning architectures show large improvement in the measured statistics $|\mu|$ and $|\Delta_{z'}>0.15|$ when the data sample is pre-cleaned using the Elliptical Envelope technique. This again shows how poorly the base routines perform on anomalous training data. As expected from the effect of outliers on the loss functions, we find that mean regression is more affected by contamination than median regression. We note that the dispersion measures $\sigma_{68}$ and $\sigma_{95}$ are very well controlled for the median regression architecture. We show the absolute values for each of the measured metrics in the third and fourth rows of Table. \ref{photoRapter}. We again show the values of each of the measured statistics, averaged over the 250 samples, for a chosen value of the contamination hyper-parameter $n_c=0.07$. 

Comparing quantile and median regression with the SOMz and Adaboost routines is not the primary focus of this work \citep[see e.g.,][]{Dietterich:2000:ECT:350128.350131,Caruana05anempirical} but we note that Adaboost with decision trees for regression is the best performing machine learning architecture on all measured statistics. However the continued success of Adaboost with contaminated data appears to be in disagreement with studies that include a large fraction of label noise in classification tasks \citep[][]{Dietterich:2000:ECT:350128.350131}. This may be an artifact of the chosen datasets and how noise is added to the data. 

\subsubsection{Anomaly detection using a cleaner galaxy sample}
\label{cleanGals}
In the previous sections we use data samples with very relaxed selection criteria, which allows both photometric, and spectroscopic data with large measured errors to be included in the base sample. We now examine the effect on the machine learning redshift if one chooses to use a base galaxy sample which has much more stringent limits of the allowed magnitude of both photometric and spectroscopic errors. We again select galaxies which pass the selection criteria described in \S\ref{estconfrac}.

We repeat the above analysis by again contaminating the base sample and using the Elliptical Envelope method to clean the sample, and then train Adaboost and SOMz for redshift analysis on both contaminated, and cleaned samples.  We again find a similar distribution of improvements in the redshift metrics as a function of the contamination hyper-parameter $n_c$, but with a slightly reduced amplitude. The improvement for Adaboost ranges from 15\% for the outlier fraction, to 85\% for the median value, and the improvement for SOMz ranges from 40\% for $\sigma_{68}$, to 95\% for the median value.

\subsubsection{Anomaly detection of non-contaminated galaxies}
We also examine the effect on the machine learning redshift if one uses only the base galaxies with a reliable spectroscopic redshift, without the addition of galaxies with unreliable redshifts. We continue as before by determining inliers and outliers as a function of the hyper-parameter $n_c$. In this section `anomalous data' could mean that a photometric magnitude in a particular band is very different from other similar galaxies at that redshift. 

\begin{figure*}
   \centering
   \includegraphics[scale=0.47,clip=true,trim=0 15 50 30]{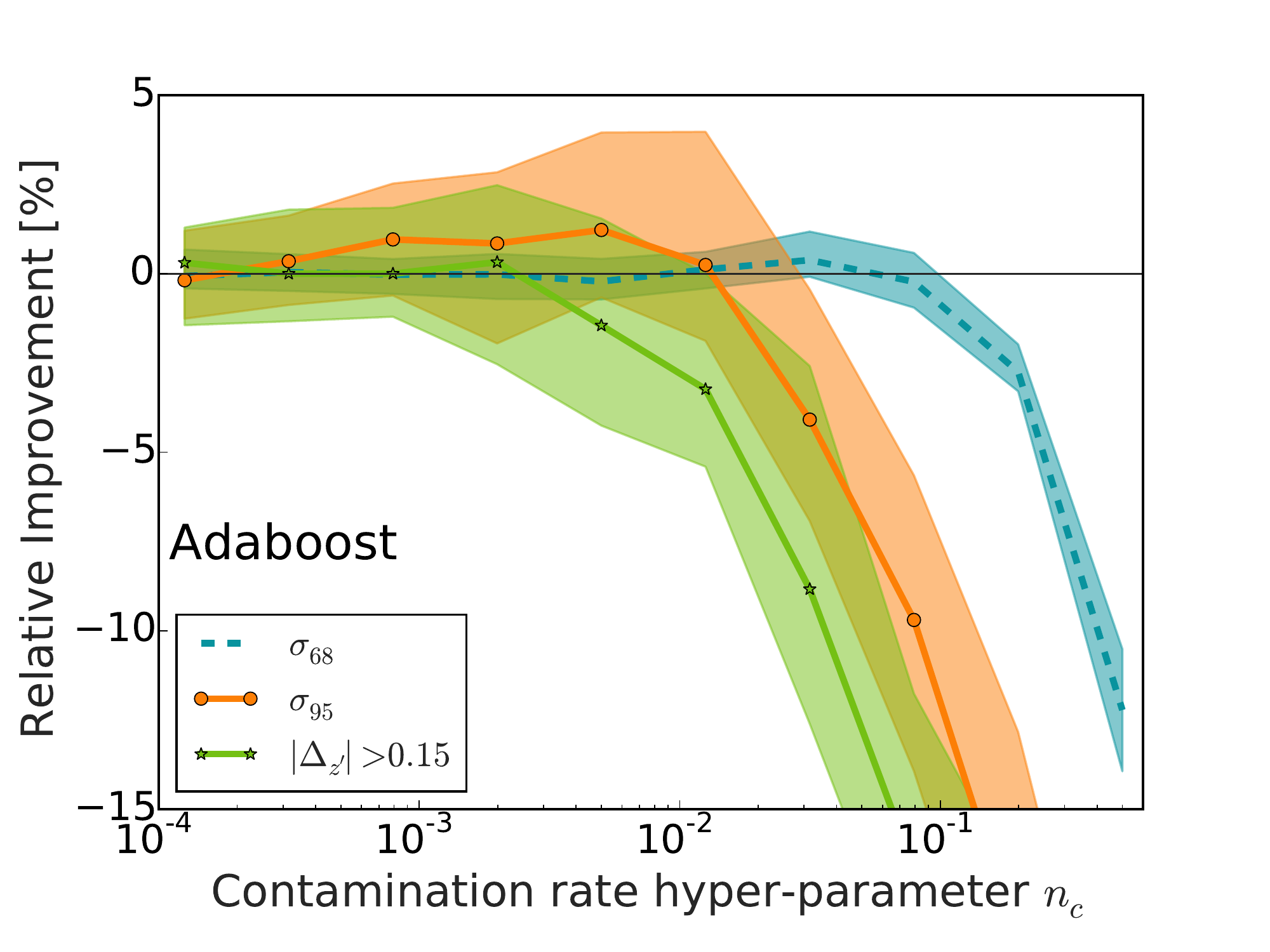}
   \includegraphics[scale=0.47,clip=true,trim=0 15 50 30]{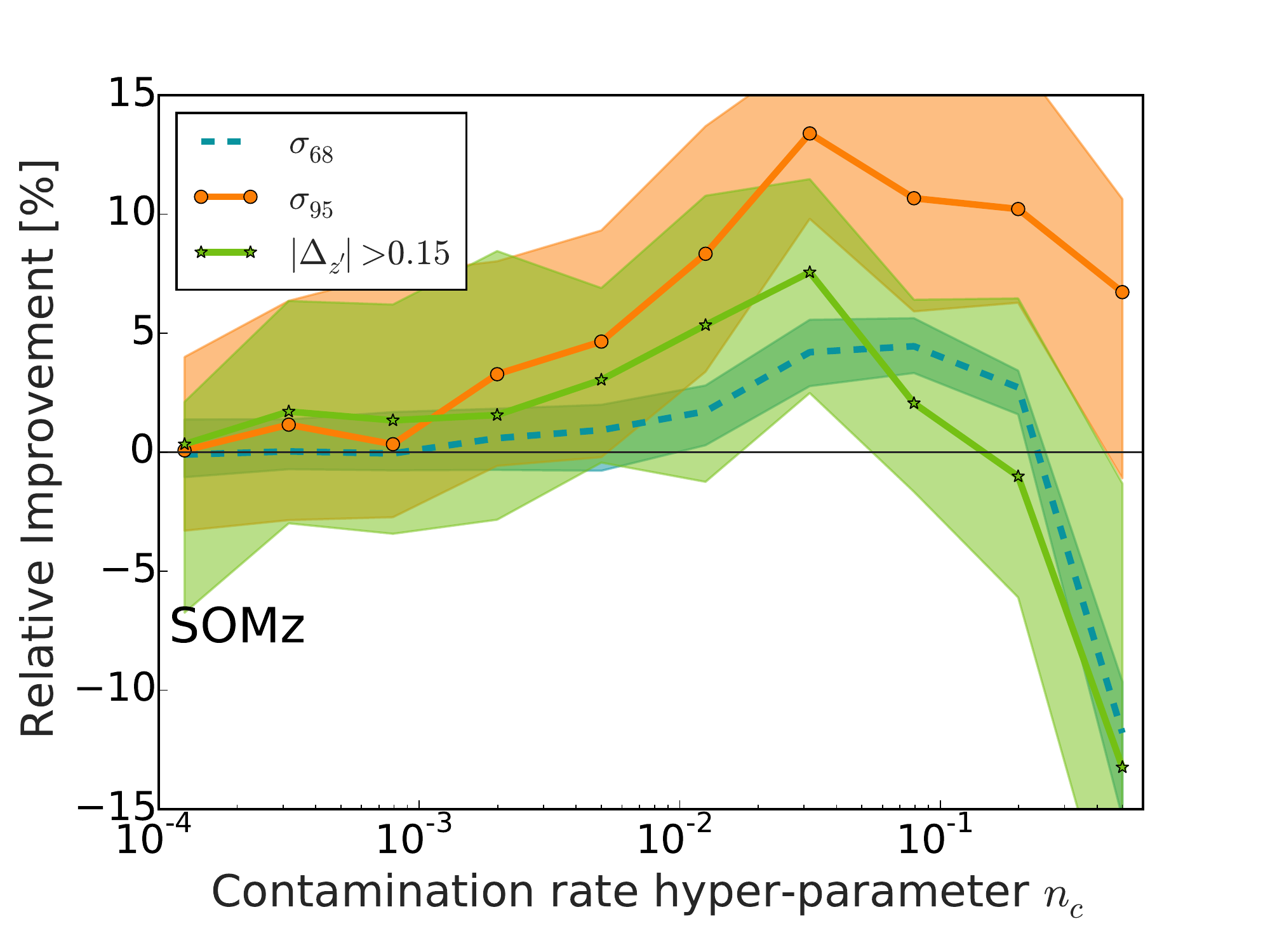}
   \caption{ \label{relativeImprovementClean} These panels are similar to those in Fig. \ref{relativeImprovement}, but with an initial data sample that does not contain galaxies with unreliable redshift estimates. Note that the y-axis scale is different between panels, and we have not shown $|\mu|$. } 
\end{figure*}

We proceed by again separately training Adaboost and the SOMz on both the initial training set and the cleaned training set. We present the results of this analysis in Fig. \ref{relativeImprovementClean}. Note that the y-axis scale is different between panels, and we have not shown $|\mu|$ due to the large scatter seen on this metric, caused by $|\mu|$ being very small.

If we adopt a contamination fraction hyper-parameter of $n_c<0.005$ and remove anomalous data, we find a very slight improvement at the level of $\sim1$\% using Adaboost and up to 4\% using SOMz in the measured metrics. Note that the relative error on $|\mu|$ is unstable, although $|\mu|$ does remain small. As $n_c$ increases, the SOMz continue to benefit from a cleaned training sample, whereas Adaboost begins to degrade in its predictive power. 

The degradation in the measured statistics seen at large values of $n_c$ in Fig. \ref{relativeImprovementClean} can be attributed to the removal of representative training data as seen in Fig. \ref{distClean}. Recall that the validation set is a random sample from the uncontaminated data with reliable redshifts, and thus would more closely resemble the solid lines in Fig. \ref{distClean}. For increasing values of $n_c$, the training and validation samples become more unrepresentative and a machine learning system would naturally degrade. We do note that SOMz appear to be less affected by small differences in the training and test data sets, but also degrade in predictive ability once the samples become very unrepresentative.

In the last two rows of Table \ref{photoRapter} we quote the median values on each of the measured statistics from each of the samples when both training on, and further cleaning, these uncontaminated galaxy samples. We note that the effect of training on the further cleaned sample improves the measured statistics using SOMz by a few percent, but can degrade some of the measured statistics by a few percent when using the Adaboost algorithm with decision trees.  

An interesting future application which is being explored by the authors is to trim the anomalous data  and then apply data augmentation \citep[see][]{2015MNRAS.450..305H} techniques to make the training and test samples again more representative.

\begin{table*}
\begin{center}
  \begin{tabular}{ | c | c | c | c | c | c |} 
 Algorithm & Sample & $\bra |\mu| \ket$ & $\bra \sigma_{68}\ket$ &  $\bra \sigma_{95}\ket$ &  $\bra |\Delta_{z'}|>0.15\ket$\\ \hline
SOMz & Inlier\&Outlier & 0.0324 & 0.077 & 0.289 & 15.06 \%\\
 & Inlier only & 0.003 & 0.032 & 0.122 & 3.35 \% \\ \hline
Adaboost & Inlier\&Outlier & 0.0046 & 0.027 & 0.167 & 4.12 \%\\
 & Inlier only & 0.0014 & 0.024 & 0.111 & 2.98 \% \\ \hline
Median regression & Inlier\&Outlier & 0.0092 & 0.087 & 0.176 & 11.22 \%\\
 & Inlier only & 0.0034 & 0.085 & 0.172 & 8.87 \% \\ \hline
Mean regression & Inlier\&Outlier & 0.0854 & 0.086 & 0.198 & 29.18 \%\\
 & Inlier only & 0.0039 & 0.078 & 0.16 & 6.75 \% \\ \hline\hline
SOMz  & (no pre-cont.) Inlier\&Outlier & 0.0009 & 0.03 & 0.134 & 3.27 \%\\
 & (no pre-cont.) Inlier only & 0.0003 & 0.029 & 0.119 & 3.15 \% \\ \hline 
Adaboost  & (no pre-cont.) Inlier\&Outlier & 0.0002 & 0.023 & 0.1 & 2.58 \%\\
 & (no pre-cont.)  Inlier only & 0.0003 & 0.023 & 0.11 & 2.98 \% \\ \hline
  \end{tabular}
\caption{\label{photoRapter} The absolute values of the different machine learning architectures applied to both data contaminated  with unreliable redshifts and the data sample with outliers removed using the Elliptical Envelope technique. The final two rows show the results applied to the data sample without initial contamination. The measured statistics are shown in the column headings, and are measured on the redshift scaled residual distribution $\Delta_{z'}$. The quoted values are the median values at fixed $n_c=0.07$ of the 250 samples that each have a different inserted contamination fraction (for the top 4 rows only). The bottom two rows use data that is not initially contaminated, although it is also cleaned using the  Elliptical Envelope technique, and are highlighted by `(no pre-cont.)' }
\end{center}
\end{table*}

\section{conclusions}
\label{conclusions}
{\rr  Machine learning methods can be used to assign redshift estimates to photometrically selected galaxy catalogues if a representative training set with both photometric properties or `features' and spectroscopic redshifts exists. Machine learning methods require that the base training sample which is used to learn the mapping  between these quantities is representative of the final, or `test', data sample. This requires that the training sample spans a similar input photometric feature space as the test sample, and does not contain anomalous data (e.g., galaxies with incorrect spectroscopic redshifts) otherwise an incorrect mapping will be learnt. In this work we examine the ability of machine learning architectures to identify and remove such anomalous data.}

{\rr In contrast to previous work on outlier analysis which removes anomalous data {\it after} the machine learning redshift system has been trained \citep[e.g.,][]{2006ApJ...651...14S,2010MNRAS.401.1399B,2014MNRAS.442.3380C}, the method presented here identifies anomalous data {\it before} the sample is used to estimate a  redshift.  The benefit of this approach is that this pre-cleaning can then be used to define a new input feature space which is much less complex than using the post processing methods. Our method makes it easier to construct of a final sample of test galaxy. }

{\rr 
The analysis in this paper uses a base sample of 2.5M galaxies drawn from the SDSS DR12 which have reliably measured spectroscopic redshifts, and some of which also have an unreliably measured spectroscopic redshift. We construct an `anomalous data sample' by selecting galaxies that have a difference between the reliable and the unreliable redshift by more than 0.01, and proceed by assigning the unreliable redshift to the galaxy. We apply this selection because we do not expect the recovered photometric redshift to have an error below $0.01$. We contaminate a base data sample with data from the anomalous sample, and then use machine learning to identify the anomalous data. }

 {\rr We choose the Elliptical Envelope routine \citep[][]{Rousseeuw:1999:FAM:331435.331458,WICS:WICS61} as the machine learning anomaly detector algorithm. The resulting ellipse encompasses a fraction of the data which are classified as `inliers' and data outside of the ellipse are classified as `outliers` or anomalous data.  We explored an alternative machine learning architecture for anomaly detection called One Class Support Vector Machines \citep[][]{SVM} and found that the Elliptical Envelope routine is more suitable for our dataset. This implies that the high dimensional data cloud is well described by a hyper-ellipse, rather than a hyper-surface with distinct regions of reliable and unreliable data which would be analysed more favourably using Support Vector Machines. There is one hyper-parameter of the Elliptical Envelope routine which is the a priori assumed contamination fraction of the data set. We describe a method to estimate this fraction using a rule-of-thumb relation between the distributions of Mahalanobis distances and the number of feature dimensions, but note the results are not very sensitive to the actual value assumed.}

{\rr 
We show how the removal of this anomalous data improves the machine learning redshift metrics for two different groups of machine learning architectures. We choose to explore both decision tree based methods and artificial Neural Network based Self Organizing Maps. These very different architectures suggest that the results found here are generalisable, and not an artefact of the machine learning method chosen. We train the machine learning systems to estimate redshifts for a test sample separately on data from the base sample contaminated with unreliable redshift estimates, and with the cleaned base sample once anomalous data has been removed.
}

{\rr
We find improvement in the all of the explored metrics when training on the cleaned sample compared with training on the contaminated sample, when comparing each machine learning method with respect to itself. We also compare the results across machine learning architectures, and find the best redshift estimation results are found using Decision Trees boosted using the AdaBoost routine \citep[][]{Freund1997119,Drucker:1997:IRU:645526.657132}. This result has been seen before by the authors \citep[][]{2015MNRAS.449.1275H}, however in that work the results are coupled with the enhanced ability of tree methods to use many tens, or hundreds of input feature dimensions.}

{\rr The SDSS data used in this work represents an optimal dataset because it covers a similar wavelength range in the photometry and spectrometry. Many other surveys do not have this luxury. For example the Dark Energy Survey \citep{2005astro.ph.10346T} has $g,r,i,z,Y$ photometry with varying depth across the sky, and have spectra drawn from heterogeneous sources. Performing outlier detection with a heterogeneous spectroscopic sample would still be possible as long as the photometry were not varying in depth drastically, otherwise even reliable data could be flagged as anomalous. If this is not the case, one potential avenue could be to degrade the entire photometry, or large fractions of it, to a similar depth and again perform outlier detection as described in this work. Furthermore we note that the spectroscopic quality flags for the SDSS data are a good estimator of reliability. This is not always true for other datasets and spectroscopic surveys e.g., the PRIMUS dataset appears to have unreliable redshift estimates for some of the most secure redshifts provided by their quality flags \citep[][see Bonnet et al in prep]{2011ApJ...741....8C,2013ApJ...767..118C}.  However one should still perform anomaly detection, even with a less reliable data sample, or one may be learning trends from spurious data.}

{\rr In this work we have also assumed that the final test sample is not contaminated by data with unreliable spectroscopic redshifts. If such a sample could not be constructed, this would not necessarily remove the usefulness of the techniques presented in this paper.  This is because a contaminated test sample would provide a similar detrimental effect to any training sample and so they would be penalised equally. This is unless the pathological case exists in which galaxies with very similar photometry, and also similar unreliable redshifts values inhabit both the training and test samples.}

{\rr An interesting avenue of future research would be to perform outlier detection on a data sample to remove anomalous training data. This may reduce the feature parameter space such that the training sample is no longer representative of the test sample. One may then employ methods from data augmentation \citep[see][]{2015MNRAS.450..305H} which enhances the training sample using third party data, from models, simulations or other dataset to make the training sample again representative of the test sample. This would work if the augmented data sample spans a similar input feature space (i.e. has the same measured photometric properties) as the training and test samples.
}

As with all machine learning works, the results found here should be applied cautiously to new datasets. Similar analysis to that described here should be performed to check if there is indeed a problem with contaminating data. If so, then we have shown that the removal of this contaminating data can greatly improve the machine learning redshift point estimates.

\begin{appendices}
\section{CasJobs MySQL query}
We obtain observational data from the SDSS using the following MySQL query which is run in the DR12 schema:
\label{sdss_q1}
\begin{verbatim}
select s.specObjID, q.objid, q.ra,q.dec, 
s.z as specz, s.zerr as specz_err, 
q.dered_u,q.dered_g,q.dered_r,q.dered_i,q.dered_z,
q.modelMagErr_u,q.modelMagErr_g,q.modelMagErr_r,
q.modelMagErr_i,q.modelMagErr_z,
q.petroRad_r,q.petroRadErr_r,
s.sourceType as specType, q.type as photpType,
s.zWarning
into mydb.specPhotoDR12v2 from SpecObjAll as s 
join photoObjAll as q on s.bestobjid=q.objid
and q.dered_g>0 and q.dered_r>0 
and q.dered_z>0 and q.type=3
\end{verbatim}
\end{appendices}

\section*{Acknowledgments} 
\label{ack}
{\rr We thank the anonymous referee for comments and suggestions which have improved the paper.}
S.Seitz and M. M. Rau are supported by the Transregional Collaborative
Research Centre TRR 33 - The Dark Universe and the DFG
cluster of excellence ``Origin and Structure of the Universe''. CB:  Funding for this project was partially provided by the Spanish Ministerio de Economa y Com- petitividad (MINECO) under projects FPA2013-47986, and Centro de Excelencia Severo Ochoa SEV-2012-0234. Funding for the SDSS and SDSS-II has been provided by the Alfred
P. Sloan Foundation, the Participating Institutions, the
National Science Foundation, the U.S. Department of
Energy, the National Aeronautics and Space Administration,
the Japanese Monbukagakusho, the Max Planck
Society, and the Higher Education Funding Council for
England. The SDSS Web Site is http://www.sdss.org/. 
\bibliographystyle{mn2e}
\bibliography{photoz}

\begin{thebibliography}{}

\bibitem[\protect\citeauthoryear{{Alam}, {Albareti} \& et al.}{{Alam}
  et~al.}{2015}]{2015arXiv150100963A}
{Alam} S.,  {Albareti} F.~D.,    et al. 2015, ArXiv 1501.00963

\bibitem[\protect\citeauthoryear{{Bernstein} \& {Huterer}}{{Bernstein} \&
  {Huterer}}{2010}]{2010MNRAS.401.1399B}
{Bernstein} G.,  {Huterer} D.,  2010, \mnras, 401, 1399

\bibitem[\protect\citeauthoryear{{Blanton} \& {Roweis}}{{Blanton} \&
  {Roweis}}{2007}]{2007AJ....133..734B}
{Blanton} M.~R.,  {Roweis} S.,  2007, \aj, 133, 734

\bibitem[\protect\citeauthoryear{{Bonnett}}{{Bonnett}}{2015}]{2015MNRAS.449.1043B}
{Bonnett} C.,  2015, \mnras, 449, 1043

\bibitem[\protect\citeauthoryear{Breiman, Friedman, Olshen \& Stone}{Breiman
  et~al.}{1984}]{ig}
Breiman L.,  Friedman J.~H.,  Olshen R.~A.,    Stone C.~J.,  1984,
  Classification and Regression Trees.
Wadsworth International Group, Belmont, CA

\bibitem[\protect\citeauthoryear{{Carrasco Kind} \& {Brunner}}{{Carrasco Kind}
  \& {Brunner}}{2013}]{tpz}
{Carrasco Kind} M.,  {Brunner} R.~J.,  2013, \mnras, 432, 1483

\bibitem[\protect\citeauthoryear{{Carrasco Kind} \& {Brunner}}{{Carrasco Kind}
  \& {Brunner}}{2014a}]{2014MNRAS.442.3380C}
{Carrasco Kind} M.,  {Brunner} R.~J.,  2014a, \mnras, 442, 3380

\bibitem[\protect\citeauthoryear{{Carrasco Kind} \& {Brunner}}{{Carrasco Kind}
  \& {Brunner}}{2014b}]{2014MNRAS.438.3409C}
{Carrasco Kind} M.,  {Brunner} R.~J.,  2014b, \mnras, 438, 3409

\bibitem[\protect\citeauthoryear{Caruana \& Niculescu-Mizil}{Caruana \&
  Niculescu-Mizil}{2005}]{Caruana05anempirical}
Caruana R.,  Niculescu-Mizil A.,  2005, in In Proc. 23 rd Intl. Conf. Machine
  learning (ICML’06 An empirical comparison of supervised learning algorithms
  using different performance metrics.
pp 161--168

\bibitem[\protect\citeauthoryear{{Coil}, {Blanton}, {Burles}, {Cool},
  {Eisenstein}, {Moustakas}, {Wong}, {Zhu}, {Aird}, {Bernstein}, {Bolton} \&
  {Hogg}}{{Coil} et~al.}{2011}]{2011ApJ...741....8C}
{Coil} A.~L.,  {Blanton} M.~R.,  {Burles} S.~M.,  {Cool} R.~J.,  {Eisenstein}
  D.~J.,  {Moustakas} J.,  {Wong} K.~C.,  {Zhu} G.,  {Aird} J.,  {Bernstein}
  R.~A.,  {Bolton} A.~S.,    {Hogg} D.~W.,  2011, \apj, 741, 8

\bibitem[\protect\citeauthoryear{{Cool}, {Moustakas}, {Blanton}, {Burles},
  {Coil}, {Eisenstein}, {Wong}, {Zhu}, {Aird}, {Bernstein}, {Bolton}, {Hogg} \&
  {Mendez}}{{Cool} et~al.}{2013}]{2013ApJ...767..118C}
{Cool} R.~J.,  {Moustakas} J.,  {Blanton} M.~R.,  {Burles} S.~M.,  {Coil}
  A.~L.,  {Eisenstein} D.~J.,  {Wong} K.~C.,  {Zhu} G.,  {Aird} J.,
  {Bernstein} R.~A.,  {Bolton} A.~S.,  {Hogg} D.~W.,    {Mendez} A.~J.,  2013,
  \apj, 767, 118

\bibitem[\protect\citeauthoryear{Cortes \& Vapnik}{Cortes \&
  Vapnik}{1995}]{SVM}
Cortes C.,  Vapnik V.,  1995, Machine Learning, 20, 273

\bibitem[\protect\citeauthoryear{{Cunha}, {Huterer}, {Lin}, {Busha} \&
  {Wechsler}}{{Cunha} et~al.}{2014}]{2014MNRAS.444..129C}
{Cunha} C.~E.,  {Huterer} D.,  {Lin} H.,  {Busha} M.~T.,    {Wechsler} R.~H.,
  2014, \mnras, 444, 129

\bibitem[\protect\citeauthoryear{{Dahlen}}{{Dahlen}}{2013}]{2013ApJ...775...93D}
{Dahlen} T. e.~a.,  2013, \apj, 775, 93

\bibitem[\protect\citeauthoryear{Dietterich}{Dietterich}{2000}]{Dietterich:2000:ECT:350128.350131}
Dietterich T.~G.,  2000, Mach. Learn., 40, 139

\bibitem[\protect\citeauthoryear{Drucker}{Drucker}{1997}]{Drucker:1997:IRU:645526.657132}
Drucker H.,  1997, in Proceedings of the Fourteenth International Conference on
  Machine Learning ICML '97, Improving regressors using boosting techniques.
Morgan Kaufmann Publishers Inc., San Francisco, CA, USA, pp 107--115

\bibitem[\protect\citeauthoryear{{Eisenstein}
  D.~J.}{{Eisenstein}}{2011}]{2011AJ....142...72E}
{Eisenstein} D.~J. e.~a.,  2011, \aj, 142, 72

\bibitem[\protect\citeauthoryear{Freund \& Schapire}{Freund \&
  Schapire}{1997}]{Freund1997119}
Freund Y.,  Schapire R.~E.,  1997, Journal of Computer and System Sciences, 55,
  119

\bibitem[\protect\citeauthoryear{Friedman}{Friedman}{1999}]{Friedman99stochasticgradient}
Friedman J.~H.,  1999, Computational Statistics and Data Analysis, 38, 367

\bibitem[\protect\citeauthoryear{Friedman}{Friedman}{2001}]{friedman2001}
Friedman J.~H.,  2001, Ann. Statist., 29, 1189

\bibitem[\protect\citeauthoryear{Geach}{Geach}{2012}]{Geach21012012}
Geach J.~E.,  2012, Monthly Notices of the Royal Astronomical Society, 419,
  2633

\bibitem[\protect\citeauthoryear{{Gerdes}, {Sypniewski}, {McKay}, {Hao},
  {Weis}, {Wechsler} \& {Busha}}{{Gerdes} et~al.}{2010}]{2010ApJ...715..823G}
{Gerdes} D.~W.,  {Sypniewski} A.~J.,  {McKay} T.~A.,  {Hao} J.,  {Weis} M.~R.,
  {Wechsler} R.~H.,    {Busha} M.~T.,  2010, \apj, 715, 823

\bibitem[\protect\citeauthoryear{{Gunn}, {Siegmund}, {Mannery}, {Owen}, {Hull},
  {Leger}, {Carey}, {Knapp}, {York}, {Boroski}, {Kent}, {Lupton}, {Rockosi}
  et~al.,}{{Gunn} et~al.}{2006}]{Gunn:2006tw}
{Gunn} J.~E.,  {Siegmund} W.~A.,  {Mannery} E.~J.,  {Owen} R.~E.,  {Hull}
  C.~L.,  {Leger} R.~F.,  {Carey} L.~N.,  {Knapp} G.~R.,  {York} D.~G.,
  {Boroski} W.~N.,  {Kent} S.~M.,  {Lupton} R.~H.,  {Rockosi} C.~M.,    et~al.,
  2006, \aj, 131, 2332

\bibitem[\protect\citeauthoryear{Hastie, Tibshirani \& Friedman}{Hastie
  et~al.}{2001}]{hastie01statisticallearning}
Hastie T.,  Tibshirani R.,    Friedman J.,  2001, The Elements of Statistical
  Learning.
Springer Series in Statistics, Springer New York Inc., New York, NY, USA

\bibitem[\protect\citeauthoryear{{Hildebrandt}, {Arnouts}, {Capak},
  {Moustakas}, {Wolf} \& {Abdalla}}{{Hildebrandt}
  et~al.}{2010}]{2010A&A...523A..31H}
{Hildebrandt} H.,  {Arnouts} S.,  {Capak} P.,  {Moustakas} L.~A.,  {Wolf} C.,
   {Abdalla} e.~a.,  2010, \aap, 523, A31

\bibitem[\protect\citeauthoryear{{Hoyle}, {Rau}, {Bonnett}, {Seitz} \&
  {Weller}}{{Hoyle} et~al.}{2015}]{2015MNRAS.450..305H}
{Hoyle} B.,  {Rau} M.~M.,  {Bonnett} C.,  {Seitz} S.,    {Weller} J.,  2015,
  \mnras, 450, 305

\bibitem[\protect\citeauthoryear{{Hoyle}, {Rau}, {Zitlau}, {Seitz} \&
  {Weller}}{{Hoyle} et~al.}{2015}]{2015MNRAS.449.1275H}
{Hoyle} B.,  {Rau} M.~M.,  {Zitlau} R.,  {Seitz} S.,    {Weller} J.,  2015,
  \mnras, 449, 1275

\bibitem[\protect\citeauthoryear{Hubert \& Debruyne}{Hubert \&
  Debruyne}{2010}]{WICS:WICS61}
Hubert M.,  Debruyne M.,  2010, Wiley Interdisciplinary Reviews: Computational
  Statistics, 2, 36

\bibitem[\protect\citeauthoryear{Kohonen}{Kohonen}{1997}]{Kohonen:1997:SM:261082}
Kohonen T.,  ed. 1997, Self-organizing Maps.
Springer-Verlag New York, Inc., Secaucus, NJ, USA

\bibitem[\protect\citeauthoryear{{Lahav}}{{Lahav}}{1997}]{1997daa..conf...43L}
{Lahav} O.,  1997, in {Di Gesu} V.,  {Duff} M.~J.~B.,  {Heck} A.,  {Maccarone}
  M.~C.,  {Scarsi} L.,   {Zimmerman} H.~U.,  eds, Data Analysis in Astronomy
  {Artificial neural networks as a tool for galaxy classification.}.
pp 43--51

\bibitem[\protect\citeauthoryear{Li \& Thakar}{Li \&
  Thakar}{2008}]{10.1109/MCSE.2008.6}
Li N.,  Thakar A.~R.,  2008, Computing in Science and Engineering, 10, 18

\bibitem[\protect\citeauthoryear{{McQuinn} \& {White}}{{McQuinn} \&
  {White}}{2013}]{2013MNRAS.433.2857M}
{McQuinn} M.,  {White} M.,  2013, \mnras, 433, 2857

\bibitem[\protect\citeauthoryear{Pedregosa et~al.,}{Pedregosa
  et~al.}{2011}]{scikit-learn}
Pedregosa F.,  et~al., 2011, Journal of Machine Learning Research, 12, 2825

\bibitem[\protect\citeauthoryear{{Rau}, {Seitz}, {Brimioulle}, {Frank},
  {Friedrich}, {Gruen} \& {Hoyle}}{{Rau} et~al.}{2015}]{2015arXiv150308215R}
{Rau} M.~M.,  {Seitz} S.,  {Brimioulle} F.,  {Frank} E.,  {Friedrich} O.,
  {Gruen} D.,    {Hoyle} B.,  2015, ArXiv 1503.08215

\bibitem[\protect\citeauthoryear{Rousseeuw \& Driessen}{Rousseeuw \&
  Driessen}{1999}]{Rousseeuw:1999:FAM:331435.331458}
Rousseeuw P.~J.,  Driessen K.~V.,  1999, Technometrics, 41, 212

\bibitem[\protect\citeauthoryear{{S{\'a}nchez}, {Carrasco Kind}, {Lin},
  {Miquel} et~al.,}{{S{\'a}nchez} et~al.}{2014}]{2014MNRAS.445.1482S}
{S{\'a}nchez} C.,  {Carrasco Kind} M.,  {Lin} H.,  {Miquel} R.,    et~al.,
  2014, \mnras, 445, 1482

\bibitem[\protect\citeauthoryear{{Schneider}, {Knox}, {Zhan} \&
  {Connolly}}{{Schneider} et~al.}{2006}]{2006ApJ...651...14S}
{Schneider} M.,  {Knox} L.,  {Zhan} H.,    {Connolly} A.,  2006, \apj, 651, 14

\bibitem[\protect\citeauthoryear{Smith et~al.,}{Smith
  et~al.}{2002}]{Smith:2002pca}
Smith J.~A.,  et~al., 2002, \aj, 123, 2121

\bibitem[\protect\citeauthoryear{{Tagliaferri}, {Longo}, {Andreon},
  {Capozziello}, {Donalek} \& {Giordano}}{{Tagliaferri}
  et~al.}{2003}]{2003LNCS.2859..226T}
{Tagliaferri} R.,  {Longo} G.,  {Andreon} S.,  {Capozziello} S.,  {Donalek} C.,
     {Giordano} G.,  2003, Lecture Notes in Computer Science, 2859, 226

\bibitem[\protect\citeauthoryear{{The Dark Energy Survey Collaboration}}{{The
  Dark Energy Survey Collaboration}}{2005}]{2005astro.ph.10346T}
{The Dark Energy Survey Collaboration} 2005, ArXiv 0510346

\bibitem[\protect\citeauthoryear{{Yeche}, {Petitjean}, {Rich}, {Aubourg},
  {Busca}, {Hamilton}, {Le Goff}, {Paris}, {Peirani}, {Pichon}, {Rollinde} \&
  {Vargas-Magana}}{{Yeche} et~al.}{2009}]{2009arXiv0910.3770Y}
{Yeche} C.,  {Petitjean} P.,  {Rich} J.,  {Aubourg} E.,  {Busca} N.,
  {Hamilton} J.~.,  {Le Goff} J.~.,  {Paris} I.,  {Peirani} S.,  {Pichon} C.,
  {Rollinde} E.,    {Vargas-Magana} M.,  2009, ArXiv 0910.3770

\end{thebibliography}

\end{document}